\documentclass[a4paper,12pt]{article}
\pdfoutput=1
\usepackage{graphicx}
\usepackage{amsthm}
\usepackage{amsfonts}
\usepackage{amsmath,latexsym}
\usepackage{amssymb}
\usepackage{hyperref}
\usepackage{tabularx}
\newcolumntype{M}{>{$}c<{$}}
\usepackage[matrix,arrow,curve]{xy}

\numberwithin{equation}{section} \numberwithin{figure}{section}
\numberwithin{table}{section}

\def\papertitlepage{\baselineskip 3.5ex\thispagestyle{empty}}
\def\Title#1{\baselineskip 1cm \vspace{1.5cm}%
  \begin{center}{\Large\bf #1}\end{center}\vspace{0.5cm}}
\def\Authors#1{\begin{center}\renewcommand{\thefootnote}{\fnsymbol{footnote}}{\it #1}\end{center}}
\def\Abstract{\vspace{1.0cm}%
  \begin{center}{\large\bf Abstract}\end{center}}

\renewenvironment{thebibliography}{\pagebreak[3]\par\vspace{0.6em}
\begin{flushleft}{\large \bf References}\end{flushleft}
\vspace{-1.0em}

\begin{enumerate}\if@twocolumn\baselineskip=0.6em\itemsep -0.2em
\else\itemsep -0.2em\fi\labelsep 0.1em}{\end{enumerate} }

\begin{document}
{\papertitlepage \vspace*{0cm} {\hfill
\begin{minipage}{4.2cm}
CCNH-UFABC 2017\par\noindent  November, 2017
\end{minipage}}
\Title{Numerical solution of open string field theory in Schnabl gauge}
\Authors{{\sc E. Aldo Arroyo\footnote{\tt
aldo.arroyo@ufabc.edu.br}},{\sc A. Fernandes-Silva\footnote{\tt
armando.silva@aluno.ufabc.edu.br}} and {\sc R. Szitas \footnote{\tt
renato.szitas@aluno.ufabc.edu.br}}
\\
Centro de Ci\^{e}ncias Naturais e Humanas, Universidade Federal do ABC \\[-2ex]
Santo Andr\'{e}, 09210-170 S\~{a}o Paulo, SP, Brazil ${}$ }
} 

\vskip-\baselineskip
{\baselineskip .5cm \Abstract Using traditional Virasoro $L_0$
level-truncation computations, we evaluate the open bosonic string
field theory action up to level $(10,30)$. Extremizing this
level-truncated potential, we construct a numerical solution for
tachyon condensation in Schnabl gauge. We find that the energy
associated to the numerical solution overshoots the expected value
$-1$ at level $L=6$. Extrapolating the level-truncation data for
$L\leq 10$ to estimate the vacuum energies for $L > 10$, we
predict that the energy reaches a minimum value at $L \sim 12$,
and then turns back to approach $-1$ asymptotically as $L
\rightarrow \infty$. Furthermore, we analyze the tachyon vacuum
expectation value (vev), for which by extrapolating its
corresponding level-truncation data, we predict that the tachyon
vev reaches a minimum value at $L \sim 26$, and then turns back to
approach the expected analytical result as $L \rightarrow \infty$.
 }
\newpage
\setcounter{footnote}{0}
\tableofcontents

\section{Introduction}

Schnabl's solution for tachyon condensation \cite{Schnabl:2005gv}
in Witten's open bosonic string field theory \cite{Witten:1985cc}
has been a remarkable achievement which has provided an elegant
analytic proof of Sen's first conjecture
\cite{Sen:1999mh,Sen:1999xm}. Schnabl's seminal work has allowed
the development of modern analytical and numerical techniques
\cite{Okawa:2006vm,Okawa:2012ica,Fuchs:2006an,Arroyo:2011zt,Rastelli:2006ap,Ellwood:2006ba,
Okawa:2006sn,Erler:2006hw,Erler:2006ww,Schnabl:2010tb,Fuchs:2008cc,Kiermaier:2007jg,
Takahashi:2007du,Kishimoto:2011zza,Arroyo:2009ec,AldoArroyo:2011gx,Arroyo:2010sy}
which have been used to explore new analytic solutions
\cite{Takahashi:2002ez,Erler:2009uj,Murata:2011ep,Masuda:2012cj,Hata:2012cy,Masuda:2012kt,
Bonora:2011ru,Bonora:2011ri,Kiermaier:2007ba,Mertes:2016vos,Jokel:2017vlt,Arroyo:2017mpd},
and the bosonic results have been extended to the case of open
superstring field theories
\cite{Okawa:2007ri,Erler:2007xt,Arroyo:2010fq,Gorbachev:2010zz,Arefeva:2009yxe,Arroyo:2016ajg,
Arroyo:2014pua,Arroyo:2013pha,AldoArroyo:2012if,Erler:2010pr,Erler:2007rh,Erler:2013wda}.

There are two ways of writing Schnabl's analytic solution; the
first way is in terms of the Bernoulli numbers $B_n$
\cite{Schnabl:2005gv,Fuchs:2006an,Erler:2009uj},
\begin{eqnarray}
\label{intro1} \Psi &=& \sum_{n,p} f_{n,p} ({\cal L}_0 + {\cal
L}_0^\dagger)^n \tilde c_p |0\rangle  + \sum_{n,p,q} f_{n,p,q}
({\cal B}_0 + {\cal B}_0^\dagger) ({\cal L}_0 + {\cal
L}_0^\dagger)^n \tilde c_p \tilde c_q  |0\rangle \, , \\
\label{intro2} f_{n, p} &=& \frac{1-(-1)^p}{2}\frac{\pi^{-p}}{2^{n
- 2 p + 1}} \frac{1}{n!} (-1)^
      n  B_{n - p + 1} \, , \\
\label{intro3} f_{n, p, q} &=& \frac{1-(-1)^{p + q}}{2}
\frac{\pi^{-p - q}}{2^{n - 2 (p + q) + 3}} \frac{1}{n!} (-1)^{n
-q} B_{n - p - q + 2} \, ,
\end{eqnarray}
where the operators ${\cal L}_0,{\cal B}_0$ and $\tilde c_p$ are
defined in the sliver frame and are related to the worldsheet
energy-momentum tensor $T$, the $b$ and $c$ ghosts fields
respectively. A second way of writing the solution is given in
terms of wedge states with ghost insertions,
\begin{eqnarray}
\label{intro4} \Psi &=& \lim_{N \rightarrow \infty} \Big[ \psi_N-
\sum_{n=0}^{N}
\partial_n \psi_n \Big] \; , \\
\label{intro5} \psi_n &=& \frac{2}{\pi^2} U^\dag_{n+2}U_{n+2}
\big[ (\mathcal{B}_0+\mathcal{B}^\dag_0)\tilde
c(-\frac{\pi}{4}n)\tilde c(\frac{\pi}{4}n) +\frac{\pi}{2} (\tilde
c(-\frac{\pi}{4}n) + \tilde c(\frac{\pi}{4}n)) \big] | 0\rangle \,
,
\end{eqnarray}
where $\psi_N$ with $N\rightarrow\infty$ is called the phantom
term \cite{Schnabl:2005gv,Okawa:2006vm,Erler:2006hw,Erler:2006ww}.

The above equations (\ref{intro1})-(\ref{intro5}) allow us to
write the analytic solution, either in the basis of curly ${\cal
L}_0$ eigenstates or in the Virasoro $L_0$ eigenstates and those
level expansions of the solution are very useful for the numerical
evaluation of the energy. The result of the energy obtained by
means of the curly ${\cal L}_0$ level expansion is given in terms
of a divergent series which nevertheless can be resummed
numerically by means of Pad\'{e} approximants to give a good
approximation of the expected value of the D-brane tension that
agrees with Sen's first conjecture
\cite{Schnabl:2005gv,Arroyo:2009ec}. While in the case of using
the usual Virasoro $L_0$ level expansion, the resulting expression
for the energy seems to be a convergent series which approaches to
the expected value, and therefore the use of Pad\'{e} approximants
was not necessary in that case
\cite{Schnabl:2005gv,Takahashi:2007du}.

Another interesting analysis that could be performed is to derive
Schnabl's analytical solution by numerical means, namely, using a
similar strategy that has been employed in the case of numerical
solutions constructed in other gauges like the Siegel gauge
\cite{Kostelecky:1989nt,Sen:1999nx,Rastelli:2000iu,Moeller:2000xv,Gaiotto:2002wy,Taylor:2002fy},
and the so-called $a$-gauge
\cite{Asano:2006hm,Kishimoto:2009cz,Kishimoto:2009hb}. Using the
state space of Virasoro $L_0$ eigenstates, we can write a rather
generic string field $\Psi$, subject to the gauge condition ${\cal
B}_0\Psi=0$, called the Schnabl gauge. Truncating this string
field up to some given level, we can evaluate the normalized value
of the potential defined by $V(\Psi)=-S(\Psi)/T_{25}$, where $S$
is the string field theory action and $T_{25}$ represents the
value of the D-brane tension. Explicitly, the normalized potential
is given by
\begin{align}
\label{intro6} V(\Psi) = 2 \pi^2 \Big[ \frac{1}{2} \langle \Psi ,Q
\Psi \rangle + \frac{1}{3} \langle \Psi, \Psi* \Psi\rangle \Big].
\end{align}

Extremising this potential (\ref{intro6}) and keeping the
coefficient corresponding to the tachyon state fixed, we obtain
the effective tachyon potential. Actually, we will obtain many
branches for this effective tachyon potential. The configuration
corresponding to Schnabl's solution can be identified with the
local minimum of the branch which connects the perturbative with
the non-perturbative vacuum.

Using curly ${\cal L}_0$ level-truncation computations, i.e.,
working out in the sliver frame, the first attempt to obtain
Schnabl's solution numerically has been done in reference
\cite{AldoArroyo:2009hf}. In the sliver frame, the level of a
state is defined as the eigenvalue of the operator ${\cal L}_0+1$.
For instance, the truncated level one string field, following the
notation of reference \cite{AldoArroyo:2009hf}, is given by
\begin{align}
\label{intro7} \Psi = x_0 \tilde c_{1}| 0\rangle - 2 x_1
(\mathcal{L}_0+\mathcal{L}^\dag_0)\tilde c_1 | 0\rangle - 2 x_1
(\mathcal{B}_0+\mathcal{B}^\dag_0)\tilde c_0 \tilde c_1 | 0\rangle
+ (\text{higher level terms}),
\end{align}
where the coefficients of the expansion $x_0$ and $x_1$ were
chosen so that $\Psi$ satisfies the Schnabl gauge. Replacing
equation (\ref{intro7}) into equation (\ref{intro6}), we obtain
the level $V^{(1,3)}(x_0,x_1)$ potential and by integrating out
the coefficient $x_1$, namely, using $\partial_{x_1}V^{(1,3)}=0$,
we can write the coefficient $x_1$ in terms of $x_0$, to
subsequently plugging it back into the potential,
$V_{eff}^{(1,3)}(x_0)\equiv V^{(1,3)}(x_0,x_1(x_0))$, so that we
are left with the effective potential which only depends on the
coefficient $x_0$.

Since the state $\tilde c_{1}| 0\rangle$, defined in the sliver
frame, after performing the change of basis, becomes the tachyon
state $c_{1}| 0\rangle$, the effective potential that depends on
the coefficient $x_0$ has been identified as the effective tachyon
potential. However, we have noticed that the state $\tilde c_{1}|
0\rangle$ is not the only one that contains the tachyon state
$c_{1}| 0\rangle$, for instance, the truncated level four string
field contains the states $\tilde c_{-1}| 0\rangle$ and $\tilde
c_{-3}| 0\rangle$, which after performing the change of basis, it
can be shown that \cite{Schnabl:2005gv}
\begin{align}
\label{intro8} \tilde c_{-1}| 0\rangle = ( c_{-1}- c_1)| 0\rangle,
\;\;\;\;\;\;\;  \tilde c_{-3}| 0\rangle = ( c_{-3}-
\frac{1}{3}c_{-1}+\frac{1}{3}c_1)| 0\rangle.
\end{align}
Clearly these states also contain the tachyon state $c_{1}|
0\rangle$. This observation implies that the tachyon state can get
more contributions coming from states that appear at higher
levels.

Leaving aside the above subtlety, and considering a truncated
level five string field, like the one given in equation
(\ref{intro7}), it has been shown that there exist a branch of the
potential which connects the perturbative with the
non-perturbative vacuum and its local minimum occurs at a point
where $x_{0} = 0.63680186$ \cite{AldoArroyo:2009hf}. Note that,
using equation (\ref{intro2}), the analytical value of this
coefficient turns out to be $x_{0,exact}=f_{0,1} = 2/\pi
=0.63661977$. By computing the value of the remaining coefficients
and evaluating the energy, the results of \cite{AldoArroyo:2009hf}
suggest that the numerical solution, found by means of curly
${\cal L}_0$ level-truncation computations, seems to converge to
the Schnabl's analytical solution \cite{Schnabl:2005gv}.

Since the first term alone appearing in the curly ${\cal L}_0$
level expansion (\ref{intro7}) does not represent the tachyon
state, the effective potential depending on the single coefficient
$x_0$ can not be identified as being the effective tachyon
potential. In order to properly determine the effective tachyon
potential, we must use an expansion of the string field such that
the lowest state should correspond to the tachyon state alone.
This calculation can be done if we express directly the string
field in the Virasoro basis of $L_0$ eigenstates\footnote{As
usual, we define the level $L$ of a state as the eigenvalue of the
operator $L_0+1$.}. For instance, up to level two states,
following the notation of Sen and Zwiebach, we have
\begin{align}
\label{intro9} \Psi = t c_1 |0\rangle + u c_{-1} |0\rangle + v
L^{m}_{-2} c_{-1} |0\rangle + w b_{-2} c_0 c_1 |0\rangle +
(\text{higher level terms})=t \mathcal{T} +\chi,
\end{align}
where $t$, $u$, $v$ and $w$ are some unknown coefficients, and
$L^{m}_{p}$ denotes the modes of the matter Virasoro operator.

We have defined the field $\mathcal{T} \equiv c_1 |0\rangle$ as
being the tachyon contribution of the string field, while $\chi$
represents the remaining terms which are linearly independent of
the first term $\mathcal{T}$. To obtain the effective tachyon
potential, we must integrate out the string field $\chi$, this is
done by inserting the string field $\Psi$ into the potential
(\ref{intro6}), solving the equation of motion for $\chi$ and
plugging back to the action. The resulting expression, as a
function of the single variable $t$ is the effective tachyon
potential. Note that the effective tachyon potential computed in
this way is non-unique since it depends on the choice of a
specific gauge to fix the string field $\Psi$. Historically, the
gauge used has been the Siegel gauge $b_0 \Psi = 0$, and the first
numerical tests of Sen's first conjecture were done in this gauge
\cite{Kostelecky:1989nt,Sen:1999nx,Rastelli:2000iu,Moeller:2000xv,Gaiotto:2002wy}.
The most recent Virasoro $L_0$ level-truncation computations in
Siegel gauge has been performed in reference
\cite{Kishimoto:2011zza}, where the author obtained a numerical
solution up to the level $(26, 78)$.

Using Virasoro $L_0$ level-truncation computations, we are going
to derive a numerical solution $\Psi$ for tachyon condensation in
Schnabl gauge ${\cal B}_0\Psi=0$. Since the operator ${\cal B}_0$
contains all even positive modes of the $b$ ghost field
\begin{align}
\label{intro10} {\cal B}_0 = b_0 + \sum_{k=1}^{\infty}
\frac{2(-1)^{k+1}}{4k^2-1} b_{2k},
\end{align}
Schnabl gauge fixing condition turns out to be level dependent.
For instance, regarding the $w$ coefficient, if we impose Schnabl
gauge to the truncated level two string field (\ref{intro9}), we
obtain $w=0$, while using a truncated level four string field,
Schnabl gauge implies that $w=- 2E/3$, where $E$ is the
coefficient in front of the state $b_{-2} c_{-2} c_{1}|0\rangle$
which appears at level four. This result is in contrast to the
case of Siegel gauge, where the gauge condition $b_0 \Psi = 0$
implies that the coefficients satisfy some relations that are
independent of the level of the truncated string field.

In reference \cite{Schnabl:2005gv}, the author conjectured that
the level dependent Schnabl gauge fixing condition would not pose
problems and that the numerical high level computations of Moeller
and Taylor \cite{Moeller:2000xv} and Gaiotto and Rastelli
\cite{Gaiotto:2002wy} would converge to his analytical solution.
One of the main motivations for writing this paper has been to
test this conjecture by means of explicit numerical computations.

This paper is organized as follows. In section 2, by writing a
string field $\Psi$ in terms of the elements contained in the
state space of Virasoro $L_0$ eigenstates, we study and discuss
Schnabl gauge condition $\mathcal{B}_0 \Psi=0$, then using this
string field expanded up to some level $L\leq 10$, we define and
evaluate the truncated $(L,3L)$ potential. In section 3, employing
this truncated $(L,3L)$ potential, and integrating out the
non-tachyonic fields, we construct the effective tachyon potential
and analyze its branch structure. In section 4, we analyze and
extrapolate the data of the vacuum energy. In section 5, we study
and extrapolate the data of the tachyon vev. Finally, a summary
and further directions of exploration are given in the last
section.


\section{Level truncation and Schnabl gauge}
To perform level-truncation computations, first we define the
level $L$ of a state as the eigenvalue of the operator $L_0+1$.
For instance, the zero momentum tachyon state $c_1|0\rangle$ is at
level $L=0$. Let us remember that the string field action has a
twist symmetry under which all coefficients of odd-twist states
change sign, whereas coefficients of even-twist states remain
unchanged \cite{Kostelecky:1989nt,Gaberdiel:1997ia}. Therefore
coefficients of odd-twist states at levels above $c_1 |0\rangle $
must always appear in the action in pairs, and they trivially
satisfy the equations of motion if set to zero. Thus, we look for
$\Psi$ containing only even-twist states. As an example, up to
level sixth states, the truncated string field is given by
\begin{align}
\Psi = & \; t c_1 |0\rangle + u c_{-1} |0\rangle + v L^{m}_{-2}
c_{-1} |0\rangle + w b_{-2} c_0 c_1 |0\rangle + A
L^{m}_{-4} c_1|0\rangle + B L^{m}_{-2} L^{m}_{-2} c_1|0\rangle \nonumber \\
&+ C c_{-3}|0\rangle +
 D b_{-3} c_{-1} c_{1}|0\rangle + E b_{-2} c_{-2} c_{1}|0\rangle +
 F L^{m}_{-2} c_{-1}|0\rangle + w_1 L^{m}_{-3} c_0|0\rangle \nonumber \\ & +
 w_2 b_{-2} c_{-1}c_{0}|0\rangle + w_3 b_{-4} c_{0} c_{1}|0\rangle +
 w_4 L^{m}_{-2} b_{-2} c_{0} c_{1}|0\rangle +
 w_{5} c_{-5}|0\rangle + w_{6} L^{m}_{-6} c_{1}|0\rangle \nonumber \\ & + w_{7}
L^{m}_{-4} c_{-1}|0\rangle +
 w_{8} L^{m}_{-2} c_{-3}|0\rangle + w_{9} b_{-6} c_{0} c_{1}|0\rangle +
 w_{10} b_{-4} c_{-2} c_{1}|0\rangle + w_{11} b_{-4} c_{-1} c_{0}|0\rangle \nonumber \\ & +
 w_{12} b_{-2} c_{-4} c_{1}|0\rangle + w_{13} b_{-2} c_{-3} c_{0}|0\rangle +
 w_{14} b_{-2} c_{-2} c_{-1}|0\rangle + w_{15} L^{m}_{-4} L^{m}_{-2} c_{1}|0\rangle \nonumber \\ & +
 w_{16} L^{m}_{-2} L^{m}_{-2} c_{-1}|0\rangle + w_{17} L^{m}_{-4} b_{-2} c_{0} c_{1}|0\rangle +
  w_{18} L^{m}_{-2} b_{-4} c_{0} c_{1}|0\rangle +
 w_{19} L^{m}_{-2} b_{-2} c_{-2} c_{1}|0\rangle \nonumber \\ & +
 w_{20} L^{m}_{-2} b_{-2} c_{-1} c_{0}|0\rangle +
 w_{21} L^{m}_{-2} L^{m}_{-2} L^{m}_{-2} c_{1}|0\rangle +
 w_{22} L^{m}_{-2} L^{m}_{-2} b_{-2} c_{0} c_{1}|0\rangle \nonumber \\ & + w_{23} L^{m}_{-3} c_{-2}|0\rangle +
  w_{24} L^{m}_{-5} c_{0}|0\rangle + w_{25} L^{m}_{-3} L^{m}_{-2} c_{0}|0\rangle +
 w_{26} L^{m}_{-3} L^{m}_{-3} c_{1}|0\rangle \nonumber \\ & + w_{27} b_{-3} c_{-3} c_{1}|0\rangle +
 w_{28} b_{-3} c_{-2} c_{0}|0\rangle + w_{29} b_{-5} c_{-1} c_{1}|0\rangle +
 w_{30} L^{m}_{-3} b_{-3} c_{0} c_{1}|0\rangle \nonumber \\ \label{Tsf1}   & +
 w_{31} L^{m}_{-2} b_{-3} c_{-1} c_{1}|0\rangle +
 w_{32} b_{-3} b_{-2} c_{-1} c_{0} c_{1}|0\rangle +
 w_{33}  L^{m}_{-3} b_{-2} c_{-1} c_{1}|0\rangle .
\end{align}

The next step is to impose some gauge on this truncated string
field. Traditionally in $L_0$ level-truncation computations the
Siegel gauge $b_0 \Psi = 0$ has been used. Here we are going to
impose another gauge, namely, the Schnabl gauge $\mathcal{B}_0
\Psi = 0$, where
\begin{align}
\label{Tsf2} \mathcal{B}_0  = b_0 + \sum_{k=1}^{\infty} \frac{2
(-1)^{k+1}}{4 k^2-1} b_{2k}.
\end{align}
As we are going to show, after imposing Schnabl gauge condition on
the string field, the coefficients $t$, $u$, $v$, $w$, $\cdots$
that appear in the $L_0$ level expansion of the string field will
satisfy some relations.

As an explicit example, let us impose the Schnabl gauge condition
$\mathcal{B}_0 \Psi = 0$ on the truncated level 2 string field
\begin{align}
\label{Tsf3} \Psi = t c_1 |0\rangle + u c_{-1} |0\rangle + v
L^{m}_{-2} c_{-1} |0\rangle + w b_{-2} c_0 c_1 |0\rangle.
\end{align}
Since the state $|0\rangle$ has the property that $b_{n}|0\rangle
= 0$ for $n > -2$, using the commutator and anti-commutators
\begin{align}
\label{Tsf4} [b_p,L^{m}_{q}] = 0, \;\;\;\;\;\; \{b_p,b_q\} =0,
\;\;\;\;\;\; \{b_p,c_q\} = \delta_{p+q,0},
\end{align}
the computation of $\mathcal{B}_0 \Psi $, leads to
\begin{align}
\label{Tsf5} \mathcal{B}_0 \Psi  = -w b_{-2} c_{1} |0\rangle,
\end{align}
therefore the gauge condition $\mathcal{B}_0 \Psi  = 0$ implies
that
\begin{align}
\label{Tsf6} w = 0.
\end{align}

Performing similar computations, if we impose the gauge condition
$\mathcal{B}_0 \Psi  = 0$ on a truncated level 4 string field, we
get
\begin{align}
\label{Tsf7} w_i &= 0, \;\;\;\;\;\;\;\; i =1,2,3,4. \\
\label{Tsf8} w &= - \frac{2}{3}E.
\end{align}
Note that at level 2 the relation that the coefficient $w$
satisfies is given by $w=0$, while at level 4 it has a different
relation $w=- \frac{2}{3}E$. Going further, for a truncated level
6 string field, using $\mathcal{B}_0 \Psi  = 0$, we can show that
\begin{align}
\label{Tsf9} w = -\frac{2 E}{3}+\frac{2 w_{12}}{15}.
\end{align}
So in general, it turns out that the relation satisfied by the
coefficients, after imposing the gauge condition $\mathcal{B}_0
\Psi  = 0$, depends on the level of the truncated string field.
This result is in contrast to the case of Siegel gauge, where the
gauge condition $b_0 \Psi  = 0$ implies that the coefficients
satisfy some relations that are independent of the level of the
truncated string field. For instance, regarding to the coefficient
$w$, if we impose the Siegel gauge condition $b_0 \Psi = 0$ on a
truncated level 2 string field, we obtain $w=0$. Now, if we use a
truncated level 4 string field we also get $w=0$, and even for
higher levels the same relation $w=0$ holds.

In reference \cite{Schnabl:2005gv}, the author has conjectured
that this level dependent gauge fixing would not pose problems and
that the numerical high level computations of Moeller, Taylor
\cite{Moeller:2000xv}, Gaiotto and Rastelli \cite{Gaiotto:2002wy}
would converge to his analytical solution. Extrapolating our
results to higher levels, we are going to argue that the
convergence of the numerical solution to the Schnabl's analytical
solution \cite{Schnabl:2005gv}, as $L\rightarrow \infty$, will be
very slow. In this respect convergence properties of Siegel gauge
is better than Schnabl gauge.

Next, let us compute the normalized value of the tachyon potential
which is given by
\begin{align}
\label{Tsf10} V(\Psi) = 2 \pi^2 \Big[ \frac{1}{2} \langle \Psi ,Q
\Psi \rangle + \frac{1}{3} \langle \Psi, \Psi* \Psi\rangle \Big].
\end{align}
The $V^{(L,n)}$ level truncated potential is obtained by replacing
the truncated level $L$ string field into the potential
(\ref{Tsf10}) and keeping interaction terms up to the total level
$n$, note that $2L\leqslant n \leqslant 3L $. In this paper, we
consider the maximum value of $n$, namely, we are going to work
with the truncated $(L,3L)$ potential.

Although the computation of the cubic interaction term becomes
tedious at higher levels, the evaluation of the truncated $(L,3L)$
potential is straightforward. Based on conservation laws
\cite{Rastelli:2000iu}, we have written a computer code which
evaluates higher level cubic vertices. With the help of this code,
we have obtained results up to level $(10,30)$ \footnote{We would
like to mention that in order to test our code, before computing
the numerical solution in Schnabl gauge, we have derived the
numerical solution in Siegel gauge, and shown that all our results
coincide with the ones found in references
\cite{Moeller:2000xv,Gaiotto:2002wy}.}. Once we have the
potential, the next step is to impose the gauge condition and then
find the stationary point of the potential, where in the case of
Schnabl gauge when $L \rightarrow \infty$ the stationary point
should correspond to the Schnabl's analytic solution for tachyon
condensation \cite{Schnabl:2005gv}.

Another interesting computation that can be done with the
potential is the construction of the effective tachyon potential.
In order to explain the procedure for finding the effective
tachyon potential, let us first set all components of the string
field to zero except for the first coefficient $t$. This state
will be said to be of level zero. Thus, we take
\begin{align}
\label{Tsf11} \Psi = t c_1 |0\rangle.
\end{align}
Substituting (\ref{Tsf11}) into the definition (\ref{Tsf10}), we
get the level $(0,0)$ approximation to the tachyon potential,
\begin{align}
\label{Tsf12} V^{(0,0)} = 2 \pi ^2 \Big[- \frac{t^2}{2} + \frac{27
\sqrt{3} t^3}{64}\Big].
\end{align}
The local minimum of the above potential is located at
$t_0=0.456177$, and the level $(0,0)$ potential evaluated at this
point has the value of $ V^{(0,0)}(t_0)=-0.684616$.

Going to the next level, namely using the level 2 string field
(\ref{Tsf3}), and plugging it into the definition (\ref{Tsf10}),
we obtain the following level $(2,6)$ potential
\begin{align}
V^{(2,6)} = & 2 \pi ^2 \Big[ \frac{1}{2} \Big( -t^2-u^2-6 u w+13
v^2+26 v w+4 w^2\Big) \nonumber \\ & + \frac{1}{3} \Big(\frac{81
\sqrt{3} t^3}{64} + \frac{99}{64} \sqrt{3} t^2 u-\frac{585}{64}
\sqrt{3} t^2 v+\frac{19}{64} \sqrt{3} t
   u^2+\frac{\sqrt{3} u^3}{64} -\frac{715 t u v}{32 \sqrt{3}} \nonumber \\ &-\frac{9}{4} \sqrt{3} t^2 w+\frac{3}{2} \sqrt{3} t u w+\frac{7553 t v^2}{64
   \sqrt{3}}-\frac{1235 u^2 v}{576 \sqrt{3}}+\frac{83083 u v^2}{1728
   \sqrt{3}}-\frac{272363 v^3}{1728 \sqrt{3}}  \nonumber \\ \label{Tsf13} & + \frac{65 t v w}{2 \sqrt{3}}+\sqrt{3} t w^2+\frac{703 u^2 w}{108 \sqrt{3}}-\frac{65 u
   v w}{6 \sqrt{3}}-\frac{47 u w^2}{27 \sqrt{3}}-\frac{7553 v^2 w}{108
   \sqrt{3}}-\frac{65 v w^2}{9 \sqrt{3}}\Big) \Big].
\end{align}
Now we need to impose the gauge condition on the string field.
Notice that at level 2, both the Siegel gauge condition $b_0
\Psi=0$ and the Schnabl gauge condition $\mathcal{B}_0 \Psi  = 0$
imply that $w=0$. Setting $w=0$ in the potential (\ref{Tsf13}), we
obtain the level $(2,6)$ gauge fixed potential. Since the
effective tachyon potential depends on the single variable $t$
which corresponds to the tachyon coefficient, we are going to
integrate out the rest of the coefficients $u$ and $v$. Using the
partial derivatives of the potential (with $w=0$), $\partial_u
V^{(2,6)}=0$, and $\partial_v V^{(2,6)}=0$, we can write the
coefficients $u$ and $v$ in terms of $t$.

Starting at level $(2,6)$, coefficients other than the tachyon
coefficient $t$ do not appear quadratically, therefore we cannot
exactly integrate out these coefficients $u$ and $v$. We use
Newton's numerical method to find the zeros of the partial
derivatives of the potential. For a fixed value of the tachyon
coefficient $t$, there are in general many solutions of the
equations for the remaining coefficients, which correspond to
different branches. The branch structure corresponding to the
effective tachyon potential will be analyzed in the next section.

At this point, we are interested in the branch of the effective
tachyon potential connecting the perturbative with the
non-perturbative vacuum and having a minimum value which agrees
with the one predicted from Sen's first conjecture. For instance,
the local minimum of the level $(2,6)$ effective tachyon potential
corresponds to $t_0=0.544204, u_0=0.190190, v_0=0.055963$ and the
potential evaluated at these points has the value of $
V^{(2,6)}(t_0,u_0,v_0,w=0)=-0.959376$ which is about $96\%$ of the
exact answer.

The results for the tachyon vev and vacuum energy, up to level
$(10,30)$, are shown in tables \ref{TabSie1} and \ref{TabSch1}. As
we can see, our results in Siegel gauge are the same as the ones
given in references \cite{Moeller:2000xv,Gaiotto:2002wy}. Note
that, in Schnabl gauge, at level $L=6$ the energy overshoots the
predicted analytical answer of $-1$ and appears to further
decrease at higher levels. In the case of Siegel gauge, this
phenomenon happens at level $L = 14$ \cite{Gaiotto:2002wy}. As a
first impression, it seems that the level-truncation procedure is
breaking down, in the case of Schnabl gauge for $L \geq 6$, and in
the case of Siegel gauge for $L \geq 14$. Nevertheless, by using a
clever extrapolation technique to level-truncation data obtained
in Siegel gauge for $L<18$ to estimate the vacuum energies even
for $L > 18$, in reference \cite{Gaiotto:2002wy}, the authors have
shown that the results may simply indicate that the approach of
the energy to $-1$ as $L \rightarrow \infty$ is non-monotonic,
actually it is predicted that the energy reaches a minimum value
for $L \sim 27$, but then turns back to approach asymptotically
$-1$ for $L \rightarrow \infty$.

\begin{table}[ht]
\caption{$(L,3L)$ level-truncation results for the tachyon vev and
vacuum energy in Siegel gauge.} \centering
\begin{tabular}{|c|c|c|}
\hline $L$ & $c_1 |0\rangle$ & $E^{Sie}$ \\ \hline
0 & 0.456177990470 & $-$0.684616159915\\
\hline 2 & 0.544204232320 & $-$0.959376599521 \\ \hline
4 & 0.548398986499 & $-$0.987821756244\\
\hline 6 & 0.547932362586 & $-$0.995177120537 \\ \hline 8 & 0.547052407685 & $-$0.997930183378\\
\hline
10 & 0.546260900230 & $-$0.999182458475 \\
\hline
\end{tabular}
\label{TabSie1}
\end{table}

\begin{table}[ht]
\caption{$(L,3L)$ level-truncation results for the tachyon vev and
vacuum energy in Schnabl gauge.} \centering
\begin{tabular}{|c|c|c|}
\hline $L$ & $c_1 |0\rangle$  & $E^{Sch}$ \\ \hline
0 & 0.456177990470 & $-$0.684616159915 \\
\hline 2 & 0.544204232320 & $-$0.959376599521 \\ \hline
4 & 0.548938521247 & $-$0.994651904750 \\
\hline 6 & 0.548315148955 & $-$1.003983765388 \\ \hline 8 & 0.547321883647 & $-$1.007110280219 \\
\hline
10 & 0.546508411314 &  $-$1.008189759705 \\
\hline
\end{tabular}
\label{TabSch1}
\end{table}

In the case of Schnabl gauge, applying Gaiotto-Rastelli
extrapolation and an alternative technique called as Pad\'{e}
extrapolation to the level-truncation data for $L\leq 10$ to
estimate the vacuum energies even for $L > 10$, in section 4, we
are going to predict that the energy reaches a minimum value for
$L \sim 12$, and then turns back to approach asymptotically $-1$
for $L \rightarrow \infty$.

For the case of the tachyon vev data obtained in Schnabl gauge
which is given in table \ref{TabSch1}. Note that the value of the
tachyon vev gets a maximum value near level $L=4$ and then starts
to decrease. In order to reach the analytical value of $0.553465$
\cite{Schnabl:2005gv}, there should be some higher value of $L>4$
such that the tachyon vev stops decreasing and then starts
increasing to approach asymptotically the expected result. In the
case of Siegel gauge, it may also happen that there is some value
of $L>4$ such that the tachyon vev stops decreasing. Actually,
this kind of possibility has been considered in reference
\cite{Hata:2000bj}, where the analytic value $\sqrt{3}/\pi \cong
0.551328$ was conjectured for the tachyon vev. All these issues
related to the discussion of the tachyon vev will be analyzed in
section 5.

\section{Effective tachyon potential and its branch structure}

Using the $(L,3L)$ level truncated potential with $L$=2,4,6 and 8,
in this section, we are going to analyze the branch structure of
the effective tachyon potential derived in Siegel as well as in
Schnabl gauge.

\subsection{Branch structure in Siegel gauge}

As we have seen in the previous section, the truncated level $L$
string field can be written as $\Psi= t \mathcal{T} +\chi$, where
$\mathcal{T} \equiv c_1 |0\rangle$ represents the tachyonic part
and $\chi$ is the remaining non-tachyonic contribution. By
substituting this $L_0$ level expansion of $\Psi$ into the string
field action, we derive the $V^{(L,3L)}$ level truncated
potential. Note that this potential will depend on the tachyonic
coefficient $t$ as well as on the other non-tachyonic coefficients
that are contained in the $\chi$ term. In this subsection, using
the Siegel gauge condition $b_0 \Psi=0$, we are going to study the
effective tachyon potential.

To construct the effective tachyon potential which depends only on
the tachyon coefficient $t$, we must integrate out the
non-tachyonic coefficients. Since starting at level $(2,6)$,
coefficients other than the tachyon coefficient $t$ do not appear
quadratically, we cannot exactly integrate out these non-tachyonic
coefficients. Therefore, we are forced to use numerical methods to
study the effective tachyon potential. We have used Newton's
method to find the zeros of the partial derivatives of the
potential. For a fixed value of the tachyon coefficient $t$, there
are many solutions of the equations for the remaining
coefficients, which correspond to the different branches of the
effective tachyon potential.

We are interested in the branches that are close to the physical
branch, namely, the branch connecting the perturbative with the
non-perturbative vacuum which we label as branch 1 for Siegel
gauge and branch $A$ for Schnabl gauge. In the case of Siegel
gauge, we have found four roots of the level (2,6) potential that
correspond to four branches label by 1,2,3 and 4, where branch 1
precisely corresponds to the physical branch. To analyze these
branches 1, 2 and 3 at higher levels $L>2$, we have used those
roots found at level $L=2$ as initial values, while to derive
branch 4 at higher levels, it has been necessary to find the
corresponding root of the level (4,12) potential.

\begin{figure}[h]
\centering
\includegraphics[width=6.4in,height=94mm]{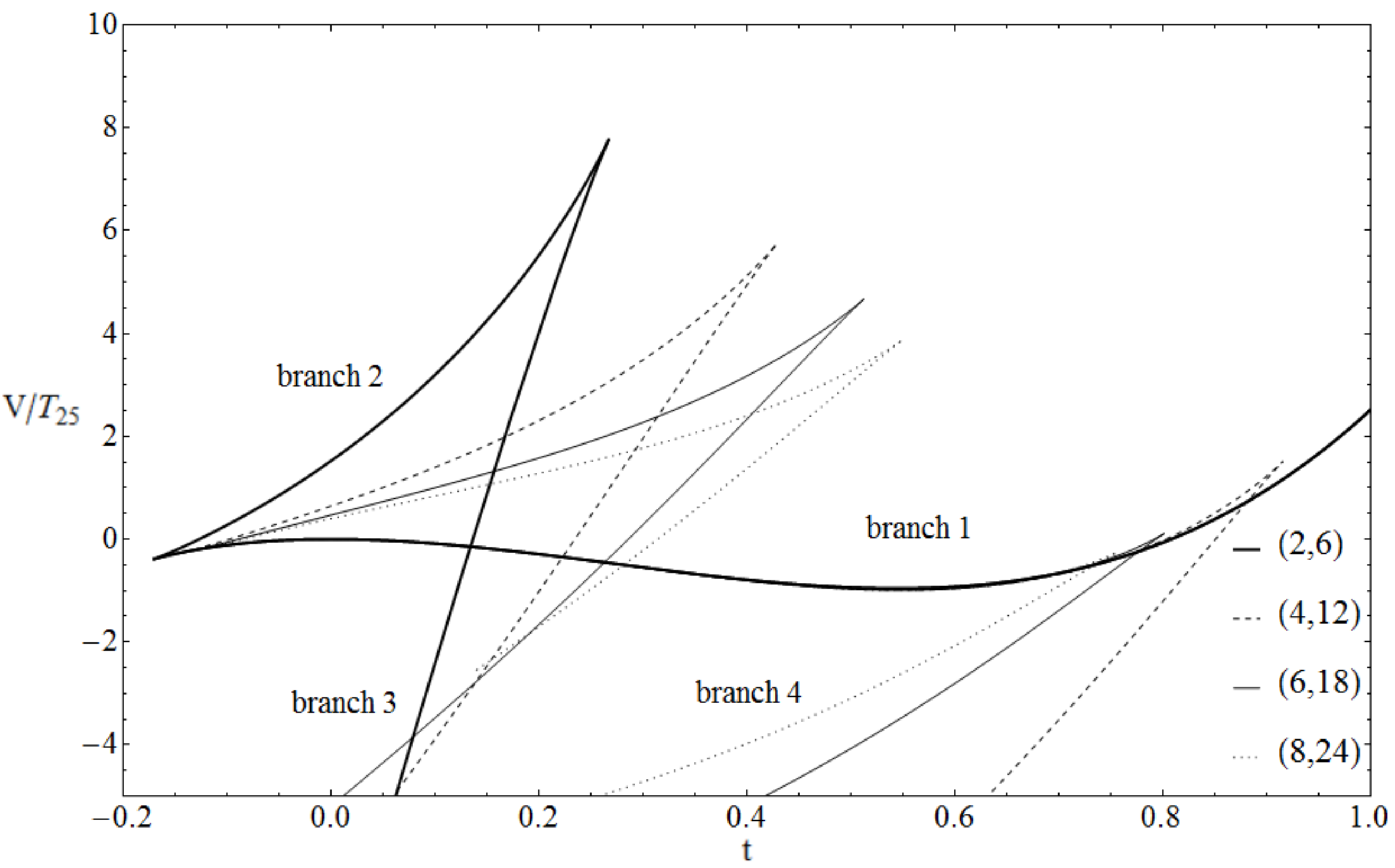}
\caption{Plot of branches 1,2,3 and 4 corresponding to the
effective tachyon potential at levels $(L,3L)$ with $L$=2,4,6 and
8, in Siegel gauge. } \label{allbranchSie}
\end{figure}

Figure \ref{allbranchSie} shows these four branches of the
effective potential at levels (2,6),(4,12),(6,18) and (8,24). The
physical branch (branch 1) has the interesting property that it
meets branches 2 and 4 at points where Newton's method becomes
unstable. The location of these points denoted as $t_{-}$ and
$t_{+}$ depends on the level and are given in table \ref{divSie}.
At level (2,6) it happens near $t_{-} \approx -0.17$ where branch
1 meets branch 2 from the left, and $t_{+} \approx 3.34$ where
branch 1 meets branch 4 from the right. As we increase the level
of the potential, we noticed that these two points converge to
some fixed values, furthermore it seems that branches 2,3 and 4
are getting closer to branch 1 in a smooth way. These properties
related to the branch structure where already discussed by Moeller
and Taylor \cite{Moeller:2000xv}, even though they have not
identified branch 4 which meets branch 1 for positive values of
the tachyon coefficient, they have mentioned the possible
existence of such a branch.

\begin{table}[h]
\centering \caption{Approximate values of the tachyon coefficients
$t$ where Newton's algorithm stops converging for branch 1 at
levels $(L,3L)$ in Siegel gauge.} \label{divSie}
\begin{tabular}{|c|c|c|c|c|}
\hline \   & $L=2$ & $L=4$ & $L=6$ & $L=8$  \\ \hline
 $t_{-}$ & $-$0.1734 & $-$0.1428 & $-$0.1336 & $-$0.1292  \\
 $t_{+}$ & \;\;\;3.3468 & \;\;\;0.9149 & \;\;\;0.8012 & \;\;\;0.7549  \\
\hline
\end{tabular}
\end{table}

\subsection{Branch structure in Schnabl gauge}

The $(L,3L)$ potential in Schnabl gauge can be constructed after
imposing the so-called Schnabl gauge condition $\mathcal{B}_0
\Psi=0$ on the string field $\Psi$. The $(L,3L)$ effective tachyon
potential is then obtained after integrating out the non-tachyonic
coefficients that appear in the expansion of $\Psi$. We have seen
that the gauge condition implies some relations that these
non-tachyonic coefficients must obey. It turns out that up to
level $L=2$, either Siegel or Schnabl gauge condition implies the
same relations for the non-tachyonic coefficients, such that up to
this level, the branch structure of the effective tachyon
potential in Schnabl gauge is exactly the same as in Siegel gauge.

Starting at level $L=4$, Schnabl gauge condition provides
relations for the non-tachyonic coefficients that are different
from the ones obtained in Siegel gauge. Therefore, in the case of
Schnabl gauge, we expect for levels $L\geq 4$ a different branch
structure for the effective tachyon potential as compared to
Siegel gauge. And in fact, we cannot extend the branches found at
level $L=2$ to higher levels (with the exception of the physical
branch). For instance, if we try to extend these branches to level
$L=4$, using as initial values for Newton's method the zeros found
from the partial derivatives of the level $(2,6)$ potential, we
discover that the algorithm converges to a single solution which
precisely corresponds to the physical branch.

Thus, to study properly the branch structure of the effective
tachyon potential in Schnabl gauge, for a fixed value of the
tachyon coefficient $t$, we have been required to obtain all the
zeros of the partial derivatives of the level $(4,12)$ potential.
As a result, we have found many different solutions (branches)
including the physical branch (named as branch $A$), most of these
solutions have energy scales that are far away from the physical
branch. For a matter of analysis, we will be interested in
branches that are close to the physical branch.

\begin{figure}[t]
\centering
\includegraphics[width=6.2in,height=95mm]{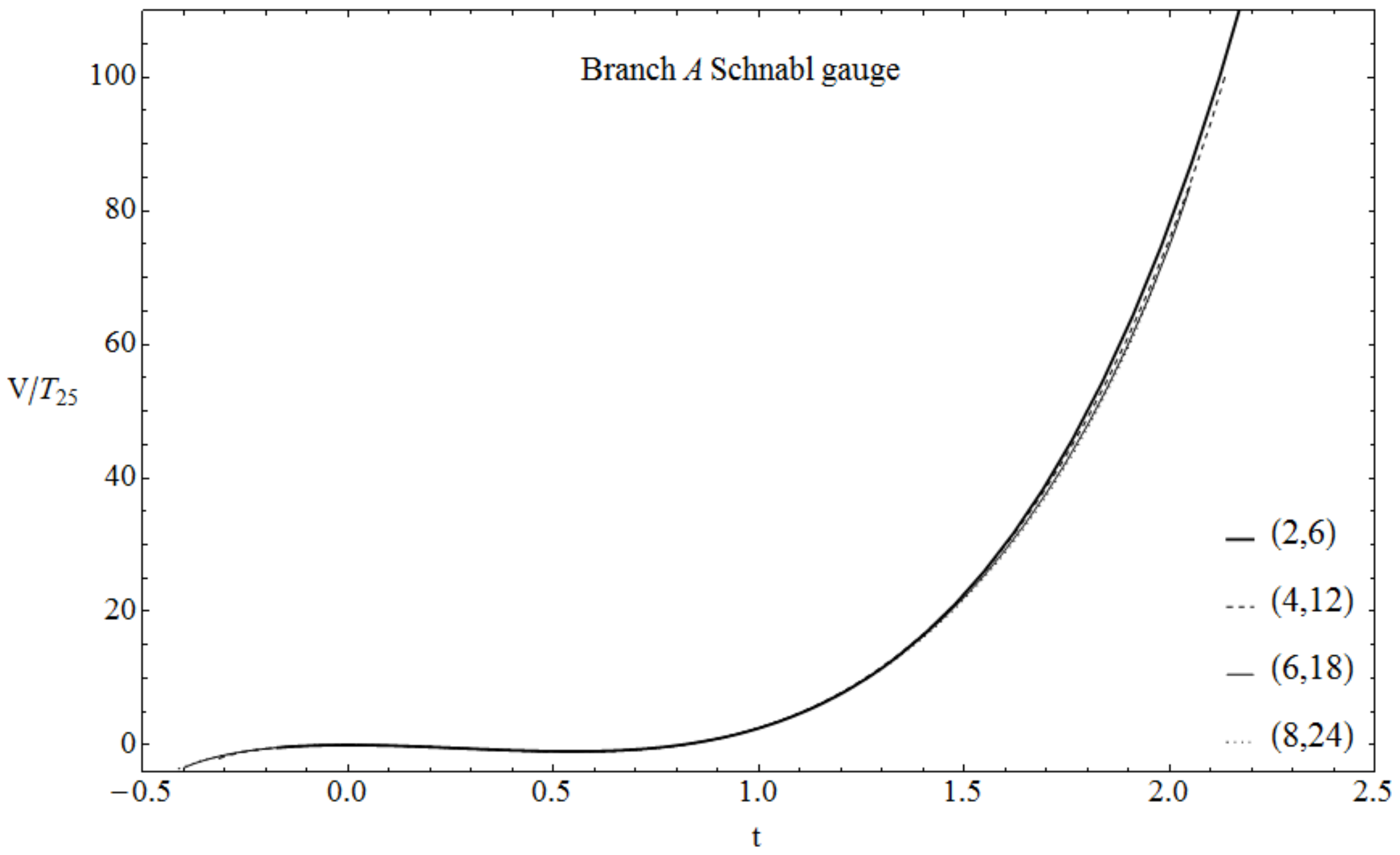}
\caption{Branch $A$ of the effective tachyon potential in Schnabl
gauge at different truncation levels.} \label{principalSch}
\end{figure}

In figure \ref{principalSch}, we show the plot of branch $A$ at
different truncation levels. Note that, each time we increase the
level, the profile of branch $A$ does not change significantly and
its shape looks quite similar to branch 1 of Siegel gauge. We have
also observed that the numerical algorithm used to construct
Branch $A$ fails to converge outside some region defined by
$t_{-}<t<t_{+}$. As shown in table \ref{divSch}, the locations of
the points $t_{-}$ and $t_{+}$ depend on the level and appear to
converge under level-truncation to fixed values. For a better
understanding of what happens near these points and verify if they
have the same origin as in the case of Siegel gauge, we are going
to study the structure of other branches that are close to branch
$A$.

\begin{figure}[t]
\centering
\includegraphics[width=6.2in,height=95mm]{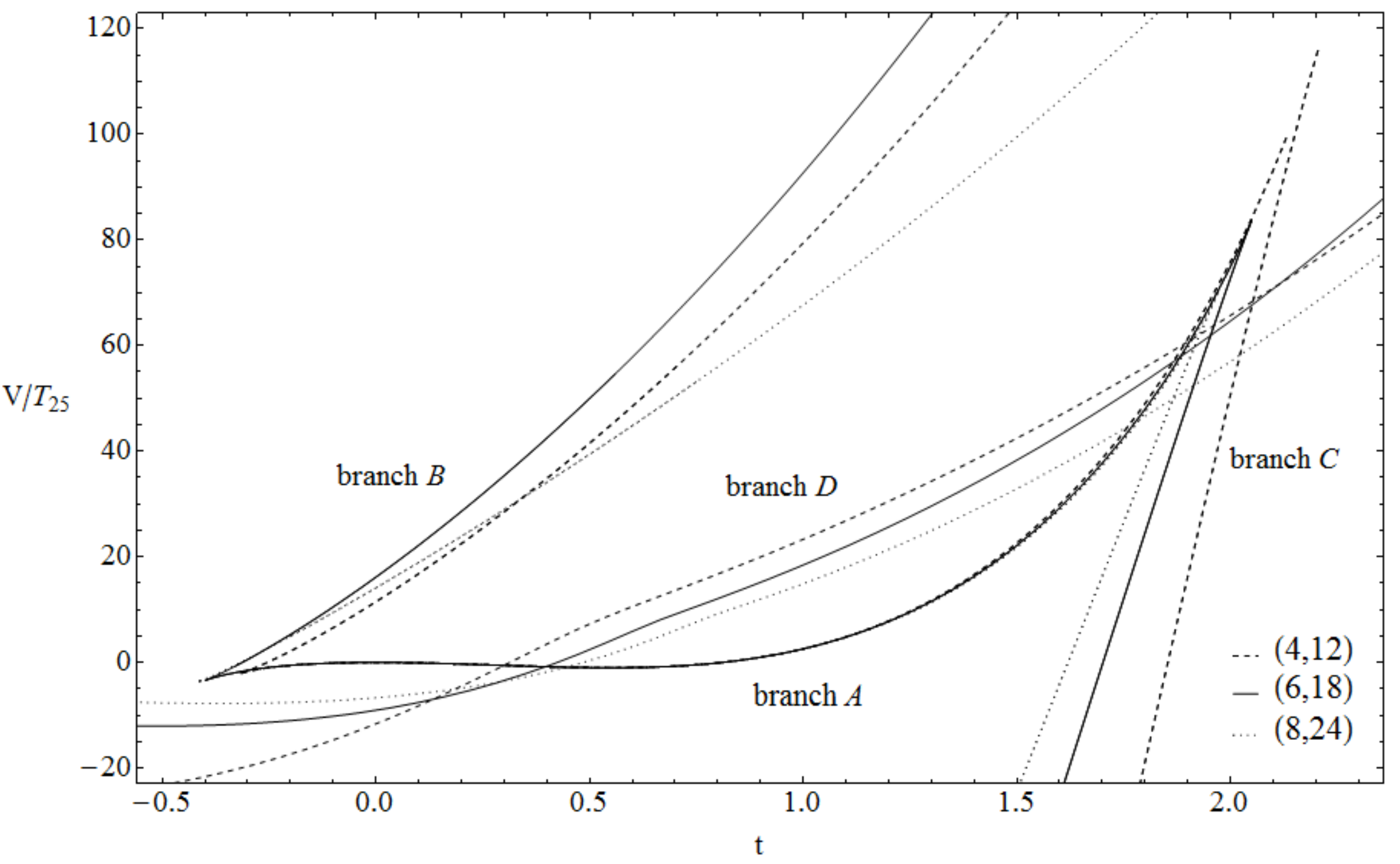}
\caption{Plot of branches $A$, $B$, $C$ and $D$ corresponding to
the effective tachyon potential at levels $(L,3L)$ with $L$=4,6
and 8, in Schnabl gauge.} \label{allbranchSch}
\end{figure}

We have discovered three different branches that are near branch
$A$, one of these branches (named as branch $D$) does not
intercept any of the other two branches at least in the region
where $t \in (t_{-},t_{+})$ and extends beyond this region of
interest. In relation to the other two branches (named as branch
$B$ and branch $C$), we observe that branch $B$ intercept branch
$A$ at the point $t_{-}$, and branch $C$ intercepts branch $A$ at
the point $t_{+}$. These branches $B$, $C$ and $D$ together with
the branch $A$ are shown in figure \ref{allbranchSch}.

Regarding the construction of branches $B$ and $C$, it turns out
that for given initial values of the non-tachyonic coefficients,
Newton's method has a limited region of convergence, therefore to
probe these branches in the region of interest defined by
$(t_{-},t_{+})$, we were required to use diverse initial values
for the non-tachyonic coefficients corresponding to different
fixed values of $t \in (t_{-},t_{+})$. We would like to point out
that branches $B$ and $C$ play similar roles as branches 2 and 4
of Siegel gauge case, namely, these branches $B$ and $C$ intercept
branch $A$ at the points where the numerical algorithm used to
construct branch $A$ becomes unstable.

\begin{table}[h]
\centering \caption{Approximate values of the tachyon coefficients
$t$ where Newton's algorithm stops converging for branch $A$ at
levels $(L,3L)$ in Schnabl gauge.} \label{divSch}
\begin{tabular}{|c|c|c|c|c|}
\hline \ & $L=4$ & $L=6$ & $L=8$ & $L=10$ \\ \hline
 $t_{-}$ & $-$0.3139 & $-$0.3992 & $-$0.4124 & $-$0.4121   \\
 $t_{+}$ & \;\;\;2.2046 & \;\;\;2.0496 & \;\;\;1.9712 & \;\;\;1.9257   \\
\hline
\end{tabular}
\end{table}

Another observation concerns how the structure of branches $B$ and
$C$ changes with the level. Branches $B$ and $C$ appear to
approach under level-truncation to branch $A$. When we move from
level (4,12) to (6,18), the slope of branch $B$ increases its
value, whereas when the level change from (6,18) to (8,24), the
slope of branch $B$ significantly decreases its value, so that
branch $B$ seems to move towards branch $A$. Regarding branch $C$,
as shown in figure \ref{allbranchSch}, the change of its slope
seems to have a smooth behavior towards branch $A$, this behavior
is similar to branch 4 of Siegel gauge. It would be interesting to
analyze the behavior of these branches at levels $(L,3L)$ with
$L>10$, however, at higher levels to obtain only a few points
along these branches should demand a lot of computing time.

Let us comment about the region $(t_{-},t_{+})$ where the
numerical algorithm used to construct branch $A$ brings convergent
results. In the case of Siegel gauge, as shown in table
\ref{divSie}, this region seems to converge approximately to
$(-0.1,0.7)$, while in the case of Schnabl gauge, it converges to
$(-0.4,1.9)$. From a mathematical point of view, it turns out that
the existence of the points $t_{-}$ and $t_{+}$ is related to the
presence of other branches which intercept the physical branch at
these points.

Regarding the physical interpretation of these branch points, let
us mention that in background independent string field theory, the
tachyon potential can be shown to have the form $V(T) =
(1+T)e^{-T}$
\cite{Gerasimov:2000zp,Kutasov:2000qp,Ghoshal:2000gt}. While the
field $T$ is related to $t$ through a nontrivial field
redefinition, it is clear that the potential $V$ is unbounded
below as $T\rightarrow -\infty$, and contains no branch points to
the left of the stable vacuum. Thus, the branch point found for
negative values of $t$ is not physical. And in fact, in the case
of Siegel gauge, there is a strong evidence that the two branch
points $t_{-}$ and $t_{+}$ appearing in the numerical computation
of the effective tachyon potential are gauge artifacts arising
when the field configuration along the effective potential leaves
the region of validity of the gauge condition
\cite{Ellwood:2001ne}. If the analysis of reference
\cite{Ellwood:2001ne} can be extended to generic gauges, our
results suggest that the region of validity of Schnabl gauge is
bigger than the region of validity of Siegel gauge.

\section{Extrapolation techniques and the vacuum energy analysis}
Using direct $(L,3L)$ level-truncation computations, we have
constructed the effective tachyon potential in Schnabl gauge up to
level $L=10$. Selecting the branch of the potential that connects
the perturbative with the non-perturbative vacuum and computing
its local minimum, we have determined the vacuum energy. As we can
see from table \ref{TabSch1}, starting at level $L=6$, the data
for the energy overshoots the conjectured value of the normalized
brane tension.

According to the data up to level $L=10$, it seems that the energy
will continue to decrease. We would like to know if this pattern
will be preserved at higher levels, namely, for levels $L>10$ the
energy continues to decrease, or as in the case of Siegel gauge
\cite{Gaiotto:2002wy} it may happen that at some level
$L_{min}>10$ the energy reaches a minimum value and then starts
increasing to approach asymptotically, as $L\rightarrow \infty$,
the expected value of $-1$. These issues could be answered if, of
course, we would have available data for levels $L>10$. Although
we do not have this data, by extrapolating the known results we
already have up to level $L=10$, we can predict the values of the
energy for levels $L>10$ which should correspond (with a good
degree of approximation) to the values obtained by means of messy
direct $(L,3L)$ level-truncation computations.

A clever extrapolation method, which we refer as Gaiotto-Rastelli
extrapolation technique, has been proposed in reference
\cite{Gaiotto:2002wy}. In the case of Siegel gauge, this technique
has been successfully used to predict the values of the energy for
levels $L>18$. In this section, after a brief review of
Gaiotto-Rastelli technique, we are going to analyze our known data
for the energy in Schnabl gauge obtained up to level $L=10$. Then,
we will study another extrapolation method, and since the function
that will be used to interpolate the known values of the energy
will be a rational function in $L$, this method will be called as
Pad\'{e} extrapolation technique.

\subsection{Gaiotto-Rastelli extrapolation technique}
In the case of Siegel gauge, in reference \cite{Gaiotto:2002wy}
using direct $(L,3L)$ level-truncation computations, the values of
the energy up to level $L=18$ were obtained, and it has been shown
that at level $L = 14$ the data overshoots the expected value of
$-1$. As a first impression, it seems that the level-truncation
procedure is breaking down for $L \geq 14$. Nevertheless,
employing a clever extrapolation technique to level-truncation
data for $L\leq18$ to estimate the vacuum energies even for $L >
18$, Gaiotto and Rastelli have shown that the results may simply
indicate that the approach of the energy to $-1$ as $L \rightarrow
\infty$ is non-monotonic, and actually it is predicted that the
energy reaches a minimum value for $L \sim 27$, but then turns
back to approach asymptotically $-1$ for $L \rightarrow \infty$.

Gaiotto-Rastelli extrapolation technique used the information of
the effective tachyon potentials $V_L(t)$. We have determined
these potentials up to level $L=10$ in Siegel as well as in
Schnabl gauges by means of $(L,3L)$ level-truncation computations.
A detailed discussion related to the effective tachyon potential
has been given in section 3, where we have analyzed the branch
structure of this potential. In this section, we are going to work
with the physical branch\footnote{Remember that this physical
branch has been labelled as branch 1, and branch $A$, for Siegel,
and Schnabl gauge respectively.}, namely, the branch that connects
the perturbative with the non-perturbative vacuum. In the case of
Schnabl gauge, this physical branch is important because when
$L\rightarrow\infty$ its local minimum should correspond to the
analytical solution found by Schnabl \cite{Schnabl:2005gv}.

Given the effective tachyon potentials derived in some gauge
$V_i(t)$ with $i=0,2,\cdots,L$, the interpolating potential is
defined as
\begin{align}
\label{GRequation} V^M_L(t)=\sum^{M/2}_{n=0}
\frac{a_n(t)}{(L+1)^n},
\end{align}
where $M$ indicates the degree of the interpolation. As we can
see, the value $1+M/2$ is equal to the number of effective
potentials contained in the set $\{V_0(t), V_2(t), V_4(t),...,
V_{L}(t)\}$ and the functions $a_n(t)$ can be expressed as linear
combinations of these potentials $V_{i}(t)$.

Before constructing the interpolating potentials $V^M_L(t)$ in
Schnabl gauge, using the effective potentials in Siegel gauge
$V^{Sie}_{i}(t)$, we are going to construct $V^M_L(t)$. We
consider first the case of Siegel gauge, since in order to test
and validate our computations, we would like to compare our
results with the well known results obtained in reference
\cite{Gaiotto:2002wy}.

\subsubsection{Siegel gauge}
As a pedagogical illustration, let us construct $V^M_L(t)$ with
$M=4$ in Siegel gauge
\begin{align}
\label{VM4Sie} V^4_L(t)=\sum^{2}_{n=0} \frac{a_n(t)}{(L+1)^n}.
\end{align}
Note that for this value of $M$, we need $1+M/2=3$ entries, that
is, the first three effective potentials (up to level $L=4$):
$\{V^{Sie}_0(t)$, $V^{Sie}_2(t)$, $V^{Sie}_4(t)\}$. To obtain the
coefficients $a_0(t)$, $a_1(t)$ and $a_2(t)$, we evaluate the
potential $V^4_L(t)$ defined in (\ref{VM4Sie}) at $L=0,2,4$ and
equate the result to the known effective potentials, namely
\begin{align}
V^4_0(t) &= V^{Sie}_{0}(t), \\
V^4_2(t) &= V^{Sie}_{2}(t), \\
V^4_4(t) &= V^{Sie}_{4}(t).
\end{align}

Solving the above system of equations, we can obtain the $a_n(t)$
coefficients as linear combinations of the potentials
$V^{Sie}_{i}(t)$
\begin{align}
\label{a04Sie} a_0(t) &= \frac{1}{8} \big[V_{0}^{Sie}(t) - 18 V_{2}^{Sie}(t) + 25 V_{4}^{Sie}(t)\big], \\
\label{a14Sie} a_1(t) &= -V_{0}^{Sie}(t) + \frac{1}{2}\big[27 V_{2}^{Sie}(t)-25 V_{4}^{Sie}(t)\big], \\
\label{a24Sie} a_2(t) &= \frac{1}{8}\big[15 V_{0}^{Sie}(t) - 90
V_{2}^{Sie}(t) + 75 V_{4}^{Sie}(t)\big]
\end{align}

Plugging these coefficients (\ref{a04Sie})-(\ref{a24Sie}) into
equation (\ref{VM4Sie}), we obtain the interpolating potential
$V^4_L(t)$ in Siegel gauge
\begin{align}
\label{VM4SieR} V^4_L(t)= \frac{\left(L^2-6 L+8\right)
V_{0}^{Sie}(t)+L (25 (L-2)V_{4}^{Sie}(t)-18 (L-4)
   V_{2}^{Sie}(t))}{8 (L+1)^2}.
\end{align}

We can use the above potential $V^4_L(t)$ to extrapolate the
values of the energy for levels $L>4$. For instance, at level $ L
= 6 $, the minimum value of $V^4_6(t)$ happens at the point where
$t_0=0.548497$ and the value of $V^4_6(t)$ evaluated at this point
gives
\begin{equation}
\label{VM4Sie6min} V^4_6(t_0)= -0.995462.
\end{equation}
This result (\ref{VM4Sie6min}) exactly matches the result found in
reference \cite{Gaiotto:2002wy}. Note that the direct $(L,3L)$
level-truncation computation (with $L=6$) brings the value of
$-0.995177$ for the vacuum energy in Siegel gauge.

By following the same procedures shown above, we can obtain the
interpolating potentials $V^M_L(t)$ for $M=6,8,10$. It turns out
that our results are identical to the results presented in
reference \cite{Gaiotto:2002wy}. Once we have learned the method
of Gaiotto-Rastelli extrapolation technique and validated our
results in Siegel gauge, we move to the case of interest, namely,
Schnabl gauge.

\subsubsection{Schnabl gauge}
To construct the interpolating potentials $V^M_L(t)$ in Schnabl
gauge, we follow the procedures explained in the case of Siegel
gauge. As a matter of illustration, the interpolating potential
$V^4_L(t)$ looks similar to the one obtained in Siegel gauge
(\ref{VM4SieR}). Actually we only need to perform the replacement
$V_{L}^{Sie}(t) \rightarrow V_{L}^{Sch}(t)$, so that
\begin{align}
\label{VM4SchR} V^4_L(t)= \frac{\left(L^2-6 L+8\right)
V_{0}^{Sch}(t)+L (25 (L-2)V_{4}^{Sch}(t)-18 (L-4)
   V_{2}^{Sch}(t))}{8 (L+1)^2}.
\end{align}
By computing the local minimum of the potential (\ref{VM4SchR})
for values of $L>4$, we can determine the extrapolated values of
the energy. For instance, at level $L = 6$, the minimum value of
$V^4_6(t)$ happens at the point where $t_0=0.549303$, and
evaluating $V^4_6(t)$ at this point gives the prediction
$V^4_6(t_0)=E^4(6) =-1.005920$ for the energy at level 6. The
direct $(L,3L)$ level-truncation computation (with $L=6$) brings
the value of $E(6)=-1.003983$.

Since we know the effective potentials in Schnabl gauge
$\{V^{Sch}_0(t), V^{Sch}_2(t), \cdots , V^{Sch}_{10}(t)\}$ up to
level $L=10$, we have determined the interpolating potentials
$V^M_L(t)$ up the maximum value of $M=10$, and by computing the
local minimum of these potentials which happens at a point close
to $t_0 \sim 0.54$, we can predict the values of the energy
$V^M_L(t_0)=E^M(L)$. Some results of these computations are shown
in table \ref{extraenergyminimum}.
\begin{table}[h]
\centering \caption{Predicted values of the energy $E^M(L)$
obtained from the interpolating potentials $V^M_L(t)$ in Schnabl
gauge, at various orders of $M$ and for $L \leq 18$. In the top
half of the table, the value of the diagonal entries $E^{M=L}(L)$
coincide with the direct $(L,3L)$ level-truncation computations.}
\label{extraenergyminimum}
\begin{tabular}{|c|c|c|c|c|}
\hline \ & $L=4$ & $L=6$ & $L=8$ & $L=10$  \\ \hline
$M=4$ & $-0.$99465190 & $-1.$00592012 & $-1.$01115471 & $-1.$01410566  \\
$M=6$ & \ & $-1.$00398376 & $-1.$00753539 & $-1.$00918262  \\
$M=8$ & \ & \ & $-1.$00711028 & $-1.$00822786  \\
$M=10$ & \ & \ & \ & $-1.$00818975  \\ \hline \hline \ & $L=12$ &
$L=14$ & $L=16$ & $L=18$ \\ \hline
$M=4$ &  $-1.$01597655 & $-1.$01725962 & $-1.$01819029 & $-1.$01889426 \\
$M=6$ &  $-1.$01004504 & $-1.$01053404 & $-1.$01082671 & $-1.$01100837 \\
$M=8$ &  $-1.$00857415 & $-1.$00859518 & $-1.$00847386 & $-1.$00829192 \\
$M=10$ & $-1.$00847289 & $-1.$00841901 & $-1.$00821931 & $-1.$00796005 \\
\hline
\end{tabular}
\end{table}

As we can see from table \ref{extraenergyminimum}, the predicted
values for the energy obtained by means of these interpolating
potentials $V^M_L(t)$ are close to the values obtained by direct
$(L,3L)$ level-truncation computations. Note also that as we
increase the value of $M$, the degree of approximation improves.
For instance, using $M=8$, namely by only knowing level-truncation
results up to level 8, one can obtain the prediction
$E^8(10)=-1.008227$ for the energy at level 10, to be compared to
the value $E(10) = -1.008189$ which has been obtained by direct
$(10,30)$ level-truncation computation.

\begin{figure}[h]
\centering
\includegraphics[width=6.0in,height=90mm]{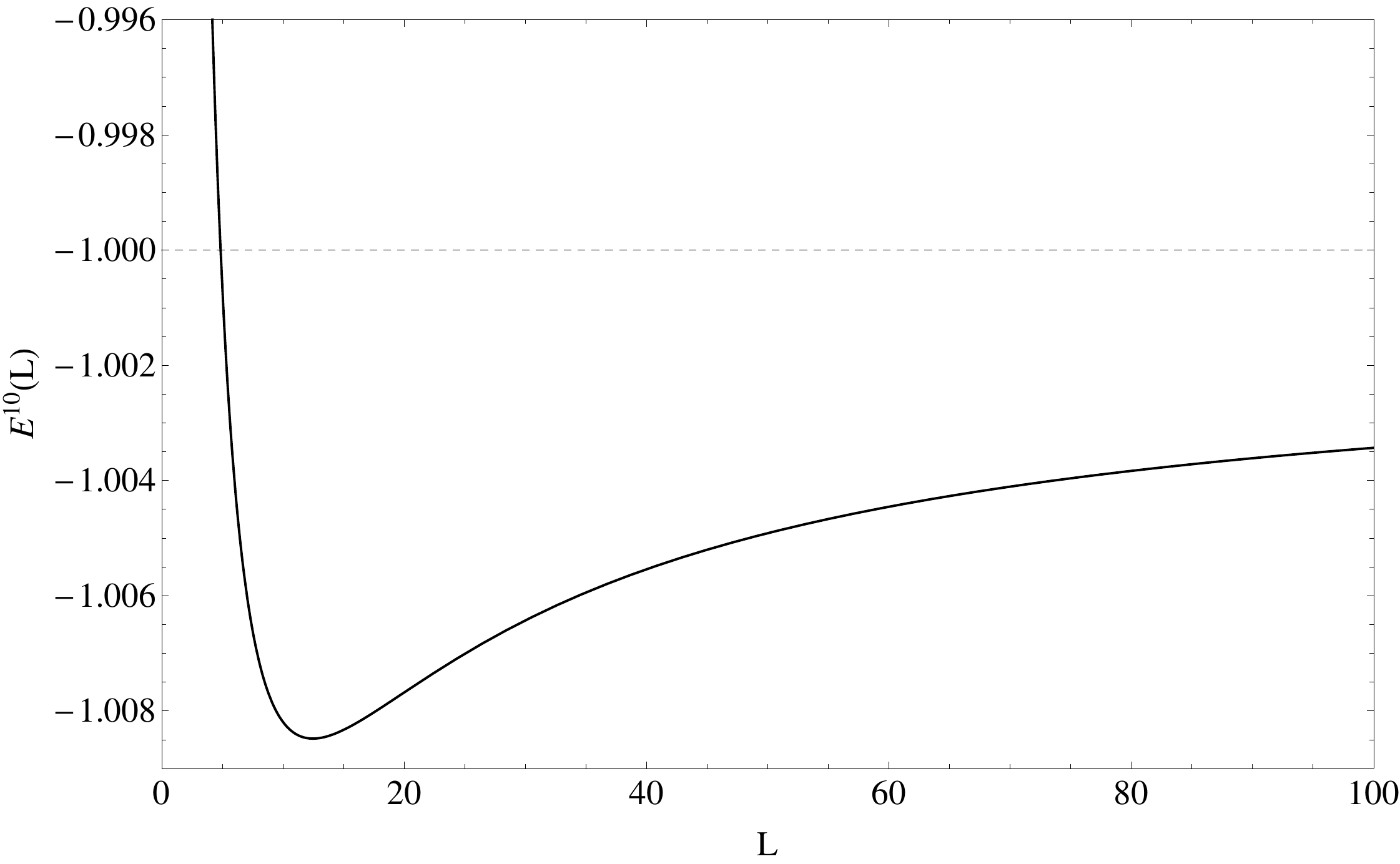}
\caption{Plot of $E^{M}(L)$ with $M=10$, for the energy in Schnabl
gauge as a function of the level. The dashed line represents the
analytical value $-1$.} \label{SchnablM10}
\end{figure}

By analyzing the predicted values of the energy $E^M(L)$ as a
function of the level, we observed that for $M>4$ the function
$E^M(L)$ behaves non-monotonically, namely, as we increase the
value of the level, the function $E^M(L)$ decreases until reaching
a minimum value $E^M(L_{min})$, then for values of the level such
that $L>L_{min}$ the function $E^M(L)$ starts increasing and
approaches asymptotically a value close to $-1$. The plot of
$E^{M}(L)$ with $M=10$, which is shown in figure \ref{SchnablM10},
illustrates clearly this point.

Using $E^M(L)$ with values of $M=6,8,10$, we have found the
corresponding values of $L_{min}$, and $E^M(L_{min})$ together
with the asymptotic one $E^M(L \rightarrow \infty)$, the results
are shown in table \ref{tableminimum}. By extrapolating the
level-truncation data for $L\leq 10$ to estimate the vacuum
energies for $L > 10$, we predict that the energy reaches a
minimum value at $L \sim 12$, and then turns back to approach
asymptotically $-1$ as $L \rightarrow \infty$. It should be nice
to confirm this prediction by means of direct $(L,3L)$
level-truncation computations.
\begin{table}[h]
\centering \caption{The local minimum $E^M(L_{min})$ and the
asymptotic $E^M(L \rightarrow \infty)$ value of $E^M(L)$ for
$M=6,8,10$ in Schnabl gauge.} \label{tableminimum}
\begin{tabular}{|c|c|c|c|}
\hline $M$ & $L_{min}$  & $E^M(L_{min})$ & $E^M(L \rightarrow
\infty)$ \\ \hline
6 & 32.98 & $-1.$00398376 & $-1.$01083980   \\
8 &  13.14 & $-1.$00711028 & $-1.$00348070   \\
10 & 12.47 & $-1.$00818977 & $-1.$00161871   \\ \hline
\end{tabular}
\end{table}

In order to bring an additional support for the previous predicted
minimum value of the energy that should happen at $L \sim 12$, in
the next subsection, we are going to use the well known results in
Siegel gauge to test another method of extrapolation technique
which will then be applied to the case of Schnabl gauge.

\subsection{Pad\'{e} extrapolation technique}
Instead of using the set of effective tachyon potentials
$V_{i}(t)$ with $i=0,2,\cdots,L$, we are going to use the values
of the local minimum of these potentials, namely, the values of
the vacuum energy which up to level $L=10$ are given in tables
\ref{TabSie1} and \ref{TabSch1} for Siegel and Schnabl gauges
respectively.

The derivation of the vacuum energy by means of $(L,3L)$
level-truncation computations is computationally less cumbersome
than the construction of the corresponding effective tachyon
potential. So in this sense, an extrapolation method that uses the
data of the vacuum energy instead of the effective potential
should be much simpler.

Given the data of the vacuum energy up to some level
$E(0),E(2),\cdots,E(L)$ derived in some gauge together with the
asymptotic expected value $E(L \rightarrow \infty)=-1$, we define
the following interpolating rational function
\begin{align}
\label{siegelnN} f_N(L)=\frac{\sum_{n=0}^{N}
a_nL^n}{1+\sum_{n=1}^{N} b_nL^n},
\end{align}
where $N$ indicates the degree of the interpolation. The value
$2N+1$ is equal to the number of elements contained in the set
$\{E(0),E(2),\cdots,E(L)\}\cup \{E(L \rightarrow \infty)=-1\}$. As
we are going to show by means of an explicit example, the
coefficients $a_n$ and $b_n$ can be determined in terms of the
data points.

Using the data of the vacuum energy in Siegel gauge up to level
$L=10$, we will construct these interpolating functions $f_N(L)$.
We consider first the case of Siegel gauge, because in order to
test the Pad\'{e} extrapolation method, we should compare our
predicted results with the well known results obtained in
reference \cite{Gaiotto:2002wy}.

\subsubsection{Siegel gauge}
As an explicit example, let us construct $f_N(L)$ with $N=1$ in
Siegel gauge
\begin{align}
\label{siegeln1} f_1(L)=\frac{ a_0+a_1 L}{1+ b_1 L},
\end{align}
for this value of $N$, we need $2N+1=3$ entries, that is the first
two vacuum energies $E^{Sie}(0)$,
 $E^{Sie}(2)$ together with the asymptotic value $-1$. To obtain
the coefficients $a_0$, $a_1$ and $b_1$, we evaluate the function
$f_1(L)$ defined in (\ref{siegeln1}) at $L=0,2,\infty$ and equate
the results to the known values, namely
\begin{align}
f_1(0) &= E^{Sie}(0)=-0.68461615991569, \\
f_1(2) &= E^{Sie}(2)= -0.95937659952124, \\
f_1(L \rightarrow \infty) &= \frac{a_1}{b_1}= -1.
\end{align}
Solving the above system of equations, we can obtain the $a_0$,
$a_1$ and $b_1$ coefficients
\begin{align}
\label{ab1Sie} a_0 = -0.68461615991569,\;\;\; a_1 =
-3.38180010003359, \;\;\; b_1 = 3.38180010003359.
\end{align}

Substituting these coefficients (\ref{ab1Sie}) into equation
(\ref{siegeln1}), we obtain the interpolating function $f_1(L)$ of
order $N=1$ in Siegel gauge.

Using this function $f_1(L)$, we can extrapolate the values of the
energy for levels $L>2$. For instance, evaluating this function at
level $ L = 4 $, we have
\begin{equation}
f_1(4)=-0.97829011.
\end{equation}
The $(L,3L)$ level-truncation computation (with $L=4$) brings the
value of $-0.98782175$ for the vacuum energy in Siegel gauge. It
turns out that as we increase the value of $N$, the degree of
approximation becomes better.

Using our known data points for the vacuum energy in Siegel gauge
up to level $L=10$ which are given in table \ref{TabSie1}, by
following the same procedures shown above, we can construct the
interpolating functions $f_N(L)$ for $N=2,3$. We have evaluated
these functions at some levels, the results are presented in table
\ref{Padesiegel}.
\begin{table}[h]
\centering \caption{Pad\'{e} extrapolation values for the vacuum
energy in Siegel gauge, from $L=4$ to $L=18$. The value of the
entries where $(4N-2=L)$ coincide with the direct $(L,3L)$
level-truncation computations.} \label{Padesiegel}
\begin{tabular}{|c|c|c|c|c|}
\hline \ & $L=4$ & $L=6$ & $L=8$ & $L=10$  \\ \hline
$N=1$ & $-0.$9782901157 & $-0.$9851868492 & $-0.$9887581331 & $-0.$9909419314  \\
$N=2$ & \ & $-0.$9951771205 & $-0.$9979360018 & $-0.$9991909062  \\
$N=3$ & \ & \ & \ & $-0.$9991824585  \\ \hline \hline \ & $L=12$ &
$L=14$ & $L=16$ & $L=18$
\\ \hline
$N=1$ &  $-0.$9924153031 & $-0.$9934764189 & $-0.$9942770695 & $-0.$9949026734 \\
$N=2$ &  $-0.$9998305359 & $-1.$0001802977 & $-1.$0003796507 & $-1.$0004953576 \\
$N=3$ &  $-0.$9998209333 & $-1.$0001702762 & $-1.$0003695899 & $-1.$0004854553 \\
\hline
\end{tabular}
\end{table}

By looking at table \ref{Padesiegel}, we note that the
extrapolated values for the energy using either $f_2(L)$ or
$f_3(L)$ for $L>10$ are almost identical. Let us remark that to
construct these interpolating functions, we have only used the
results up to level 10. So all values for the energy with $L>10$
are predictions that should be compared with the direct $(L,3L)$
level-truncation computations. For instance, in table
\ref{TabSie1Rastelli}, we show the results for the energy that
have been obtained in reference \cite{Gaiotto:2002wy} by means of
direct $(L,3L)$ level-truncation computations.
\begin{table}[ht]
\caption{$(L,3L)$ level-truncation results for the vacuum energy
in Siegel gauge extracted from reference \cite{Gaiotto:2002wy}.}
\centering
\begin{tabular}{|c|c|}
\hline $L$ & $E^{Sie}$ \\ \hline
12 & $-$0.9998222\\
\hline 14 & $-$1.0001737 \\ \hline
16 &  $-$1.0003755\\
\hline 18 &  $-$1.0004937 \\
\hline
\end{tabular}
\label{TabSie1Rastelli}
\end{table}

Comparing the results for the vacuum energy given in table
\ref{Padesiegel} (for $N=2$ or $N=3$ and $L>10$) with the results
of table \ref{TabSie1Rastelli}, we conclude that the predicted
values for the energy obtained by means of Pad\'{e} extrapolation
technique are quite well. Remarkably, the results are in agreement
up to the fifth significant digit.

Analyzing these interpolating functions with $N=2$ and $N=3$, we
observed that as we increase the value of the level, the function
$f_N(L)$ decreases until reaching a minimum value $f_N(L_{min})$,
then for values of the level such that $L>L_{min}$ the function
$f_N(L)$ starts increasing and approaches asymptotically the
expected value of $-1$. In table \ref{TurnPointPadeSiegel}, we
show the values of $L_{min}$ and $f_N(L_{min})$ for $N=2,3$. This
result is also in agreement with the result of reference
\cite{Gaiotto:2002wy}, where the authors have predicted that the
vacuum energy will reach a minimum value close to level $L \sim
27$, and then for $L>27$ approaches asymptotically the value of
$-1$.
\begin{table}[h]
\centering \caption{Local minimum values of $f_N(L)$ for $N=2,3$
in Siegel gauge} \label{TurnPointPadeSiegel}
\begin{tabular}{|c|c|c|}
\hline $N$ & $L_{min}$  & $f_N(L_{min})$  \\ \hline
2 & 26.62 & $-$1.0006243552   \\
3 &  26.72 & $-$1.0006157880  \\ \hline
\end{tabular}
\end{table}

Having tested the Pad\'{e} extrapolation technique, we are going
to apply this method to the case of Schnabl gauge.

\subsubsection{Schnabl gauge}
To construct the interpolating functions $f_N(L)$, we employ the
data for the vacuum energy in Schnabl gauge, this data is given in
table \ref{TabSch1} up to level $L=10$. Remember that we also need
to use the asymptotic value $E^{Sch}(L \rightarrow \infty)=-1$.

As a matter of illustration, let us choose $N=2$. To determine the
coefficients $a_0$, $a_1$, $a_2$, $b_1$ and $b_2$ that define the
interpolating function of order $N=2$
\begin{align}
\label{schnabln2} f_2(L)=\frac{ a_0+a_1 L+a_2 L^2}{1+ b_1 L+b_2
L^2},
\end{align}
we must solve the following system of equations
\begin{align}
f_2(0) &= E^{Sch}(0)=-0.68461615991569, \\
f_2(2) &= E^{Sch}(2)= -0.95937659952124, \\
f_2(4) &= E^{Sch}(4)=-0.99465190475076, \\
f_2(6) &= E^{Sch}(6)= -1.00398376538869, \\
f_2(L \rightarrow \infty) &= \frac{a_2}{b_2}= -1.
\end{align}
This system of equations can be easily solved to determine the
corresponding values for the coefficients $a_n$ and $b_n$.
Therefore, in this way we can also construct the interpolating
function of order $N=3$.

We have evaluated these functions $f_N(L)$ at some levels, and the
results are presented in table \ref{Padeschnabl}. Note that, as in
the case of Siegel gauge, the extrapolated values for the energy
using either $f_2(L)$ or $f_3(L)$ for $L>10$ are almost identical.

\begin{table}[h]
\centering \caption{Pad\'{e} extrapolation values for the vacuum
energy in Schnabl gauge from $L=4$ to $L=18$. The value of the
entries where $(4N-2=L)$ coincide with the direct $(L,3L)$
level-truncation computations.} \label{Padeschnabl}
\begin{tabular}{|c|c|c|c|c|}
\hline \ & $L=4$ & $L=6$ & $L=8$ & $L=10$  \\ \hline
$N=1$ & $-0.$9782901157 & $-0.$9851868492 & $-0.$9887581331 & $-0.$9909419314  \\
$N=2$ & \ & $-1.$0039837654 & $-1.$0070218437 & $-1.$0079880990  \\
$N=3$ & \ & \ & \ & $-1.$0081897597  \\ \hline \hline \ & $L=12$ &
$L=14$ & $L=16$ & $L=18$
\\ \hline
$N=1$ &  $-0.$9924153031 & $-0.$9934764189 & $-0.$9942770695 & $-0.$9949026734 \\
$N=2$ &  $-1.$0081595132 & $-1.$0080037353 & $-1.$0077135987 & $-1.$0073747284 \\
$N=3$ &  $-1.$0084698491 & $-1.$0084085581 & $-1.$0081970219 & $-1.$0079220265 \\
\hline
\end{tabular}
\end{table}

Analyzing the interpolating functions $f_N(L)$ with $N=2$ and
$N=3$, we observed that as we increase the value of the level, the
function $f_N(L)$ decreases until reaching a minimum value
$f_N(L_{min})$, then for values of the level such that $L>L_{min}$
the function $f_N(L)$ starts increasing and approaches
asymptotically the expected value of $-1$. In table
\ref{TurnPointPadeSchnabl}, we show the values of $L_{min}$ and
$f_N(L_{min})$ for $N=2,3$. This result is in agreement with the
result obtained by means of Gaiotto-Rastelli extrapolation
technique, see table \ref{tableminimum}, where we have predicted
that the energy reaches a minimum value at $L \sim 12$, and then
turns back to approach asymptotically $-1$ as $L \rightarrow
\infty$.
\begin{table}[h]
\centering \caption{Local minimum values of $f_N(L)$ for $N=2,3$
in Schnabl gauge} \label{TurnPointPadeSchnabl}
\begin{tabular}{|c|c|c|}
\hline $N$ & $L_{min}$  & $f_N(L_{min})$  \\ \hline
2 & 11.76 & $-$1.0081618558   \\
3 &  12.43 & $-$1.0084765080  \\ \hline
\end{tabular}
\end{table}

As a matter of comparison of the two extrapolation methods studied
for the case of the vacuum energy in Schnabl gauge, using
$E^{10}(L)$ and $f_3(L)$ we have constructed table \ref{TabGaPa},
where some extrapolated values for the vacuum energy from $L=12$
to $L=18$ are shown. Note that these values can be directly
extracted from tables \ref{extraenergyminimum} and
\ref{Padeschnabl} respectively. We have chosen the interpolating
functions $E^{10}(L)$ and $f_3(L)$, because these are the best
estimates we have for the vacuum energy.
\begin{table}[ht]
\caption{Extrapolated values for the vacuum energy derived by
means of $E^{10}(L)$ and $f_3(L)$, from $L=12$ to $L=18$ in
Schnabl gauge.} \centering
\begin{tabular}{|c|c|c|}
\hline $L$ & $E^{10}(L)$ & $f_3(L)$ \\ \hline
12 & $-1.$00847289 & $-1.$0084698491\\
\hline 14 & $-1.$00841901 & $-1.$0084085581 \\ \hline
16 & $-1.$00821931 & $-1.$0081970219\\
\hline 18 & $-1.$00796005 & $-1.$0079220265  \\
\hline
\end{tabular}
\label{TabGaPa}
\end{table}

By explicit $(L,3L)$ level-truncation computations with $L>10$, it
would be interesting to confirm the above predicted results. Since
the minimum value for the energy data should happen at $L \sim
12$, the direct $(L,3L)$ level-truncation computations must be
performed, at least, up to level $L=14$.

\section{The tachyon vev}

Before analyzing the tachyon vev in the case of Schnabl gauge, we
are going to study the tachyon vev in the case of Siegel gauge.

\subsection{Siegel gauge}
In the case of Siegel gauge, the results for the tachyon vev
obtained from $(L,3L)$ level-truncation computations are shown in
table \ref{TabSie1}. As we can see, the value of the tachyon vev
has a maximum value near level $L=4$ and then starts to decrease.
We would like to know if this behavior (which we call scenario
$S_1$) will continue for higher values of $L$, namely, the tachyon
vev decreases for values of $L \in (4,\infty)$ and approaches
monotonically to some asymptotic value as $L \rightarrow \infty$.
Another possible behavior (which we call scenario $S_2$) is the
one where for some very large value of $L>4$, the value of the
tachyon vev stops decreasing and then starts increasing until
reaching some asymptotic value as $L \rightarrow \infty$, i.e., a
non-monotonic behavior. Since we do not have an analytic
expression for the tachyon vacuum solution in Siegel gauge, at
this point we do not know which of these two possible scenarios
$S_1$ or $S_2$ will be the right one.

Scenario $S_2$ is compatible with the claim given in reference
\cite{Hata:2000bj} where the analytic value $\sqrt{3}/\pi \cong
0.5513$ has been conjectured for the tachyon vev. However, in
reference \cite{Gaiotto:2002wy}, the asymptotic value of $0.5405$
has been predicted for the tachyon vev, and the authors have
suggested that the conjecture \cite{Hata:2000bj} for an exact
value $\sqrt{3}/\pi$ is falsified. Therefore, according to
\cite{Gaiotto:2002wy}, scenario $S_1$ should be the correct one.
Moreover, recent numerical results up to level $L =26$ seem to
confirm this scenario \cite{Kishimoto:2011zza}\footnote{Actually
reference \cite{Kishimoto:2011zza} only discusses the results for
the energy. In relation to the tachyon vev in the traditional
Siegel gauge, I. Kishimoto has kindly shared with us his data of
$(L,3L)$ level-truncation computations for levels between $L=12$
and $L=26$.}.

Here we are going to present a criterion that will allow us to
rule out one of the two scenarios. Let us start by using a
rational function in $L$ to interpolate the data for the tachyon
vev shown in table \ref{TabSie1} together with the asymptotic
point at $L \rightarrow \infty$
\begin{align}
\label{ta1}R^{(\alpha)}_{n,Sie}(L)= \frac{a_0 + a_1 L + a_2 L^2 +
a_3 L^3 + \cdots + a_n L^n}{1 + b_1 L + b_2 L^2 + b_3
L^3+\cdots+b_n L^n}.
\end{align}
The parameter $\alpha$ has two possible values, namely, $\alpha=1$
in the case of scenario $S_1$ and $\alpha=2$ in the case of
scneario $S_2$. Since we have seven data points (which include the
point at infinity), we set $n=3$. The subscript $Sie$ means that
we are working in Siegel gauge.

Let us choose the asymptotic point given by the value
$0.540500250625$ obtained from reference \cite{Gaiotto:2002wy}, we
require that this point together with the data of table
\ref{TabSie1} match the rational function (\ref{ta1}) with $n=3$
and $\alpha=1$. Thus to determine the coefficients $a_i$ and $b_i$
we simple compare $R^{(1)}_{3,Sie}$, for each value of
$L=0,2,4,\cdots,10$ and $L \rightarrow \infty$, with all the data
points. For instance, the point at infinity should be given by
\begin{align}
\label{ta2} \lim_{ L \rightarrow \infty} R^{(1)}_{3,Sie}(L) =
\frac{a_3}{b_3} = 0.540500250625.
\end{align}
In this way we get a system of seven equations for the
coefficients $a_i$ and $b_i$ which can be easily solved. Once
these coefficients are known, the next step is to analyze the
rational function $R^{(1)}_{3,Sie}(L)$.

One thing we can do is to evaluate $R^{(1)}_{3,Sie}(L)$, for
values of $L>10$ and compare the results with the actual data of
reference \cite{Kishimoto:2011zza}. The result of these
computations for values of $L$ between $12$ to $26$ are shown in
table \ref{resultsA2}. Performing similar computations, but now
for the case of the conjectured asymptotic value
\cite{Hata:2000bj}, namely, for scenario $S_2$
\begin{align}
\label{ta3} \lim_{ L \rightarrow \infty} R^{(2)}_{3,Sie}(L) =
\frac{a_3}{b_3} = \frac{\sqrt{3}}{\pi} = 0.551328895421,
\end{align}
we obtain another set of coefficients $a_i$ and $b_i$ which define
the rational function $R^{(2)}_{3,Sie}(L)$. In table
\ref{resultsA2}, we show some values of the rational function
$R^{(2)}_{3,Sie}(L)$ compared with the direct $(L, 3L)$
level-truncation computations.

\begin{table}[ht]
\caption{The rational functions $R^{(1)}_{3,Sie}(L)$ and
$R^{(2)}_{3,Sie}(L)$ compared with the direct $(L, 3L)$
level-truncation computations in Siegel gauge.} \centering
\begin{tabular}{|c|c|c|c|}
\hline $L$ & $R^{(1)}_{3,Sie}(L)$ &$R^{(2)}_{3,Sie}(L)$ &$(L, 3L)$ results  \\
\hline
12 & 0.545607760344 & 0.545609456916 &0.545608067009 \\
\hline 14 & 0.545074327209 & 0.545079892413 & 0.545075133495 \\
\hline
16 & 0.544635438085 & 0.544646969866 & 0.544636805350\\
\hline 18 & 0.544270056598 & 0.544289404544 & 0.544271966369 \\ \hline 20 & 0.543962103018 & 0.543990840810 & 0.543964497784\\
\hline
22 & 0.543699517709 & 0.543738965854 & 0.543702325407 \\
\hline 24 & 0.543473234578 & 0.543524496424 & 0.543476381413\\
\hline
26 & 0.543276370655 & 0.543340368308 & 0.543279787348 \\
\hline
\end{tabular}
\label{resultsA2}
\end{table}

Comparing the results shown in table \ref{resultsA2}, we see that
the rational interpolating function $R^{(1)}_{3,Sie}(L)$ fits
better the actual data obtained by direct $(L,3L)$
level-truncation computations, for instance, if we compare the
data starting at level $L=22$, we can observe improvement. We can
take this result as a hint to choose $R^{(1)}_{3,Sie}(L)$ instead
of $R^{(2)}_{3,Sie}(L)$, and thus scenario $S_1$ should be more
likely than $S_2$. Next, we are going to bring another argument in
favor of scenario $S_1$.

\begin{figure}[h]
\centering
\includegraphics[width=5.5in,height=85mm]{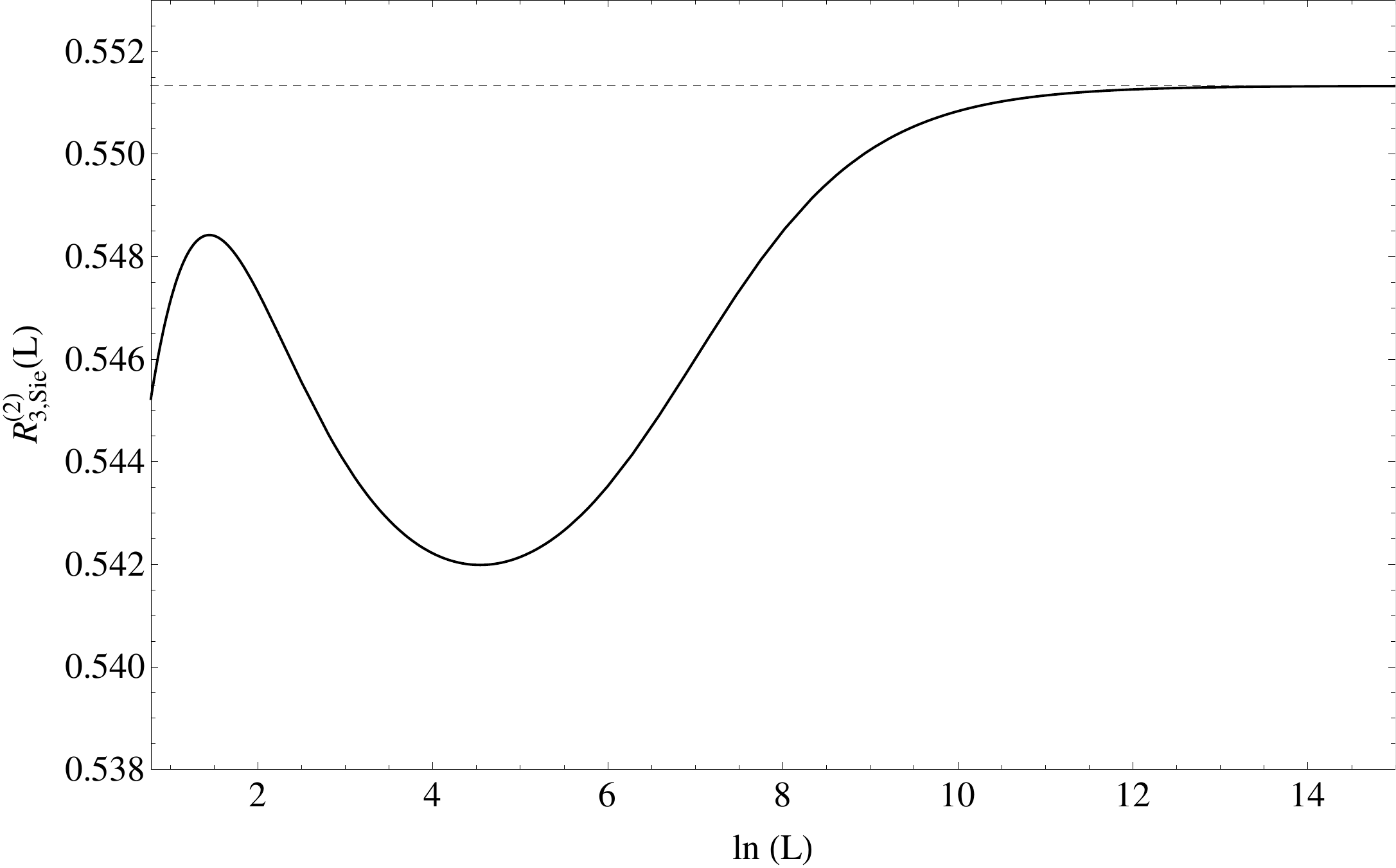}
\caption{Plot of $R^{(2)}_{3,Sie}(L)$ for the tachyon vev as a
function of level in Siegel gauge. The dashed line represents the
conjectured value $\sqrt{3}/\pi$.} \label{THfig1}
\end{figure}

For a moment, let us suppose that scenario $S_2$ is the right one,
so in this case, in order to reach the asymptotic value
$\sqrt{3}/\pi \cong 0.5513$, we would like to know at which value
of $L$ the tachyon vev starts to increase. From results up to
level $L = 26$, we can see that if there exists a point where the
tachyon vev start to increase, this value must be a very high one.
We can find this point using the interpolating rational function
$R^{(2)}_{3,Sie}(L)$. In figure \ref{THfig1}, we show a plot of
$R^{(2)}_{3,Sie}(L)$, and we can determine that the tachyon vev
starts to increase at a point close to $L \sim 94$.

\begin{figure}[h]
\centering
\includegraphics[width=5.5in,height=85mm]{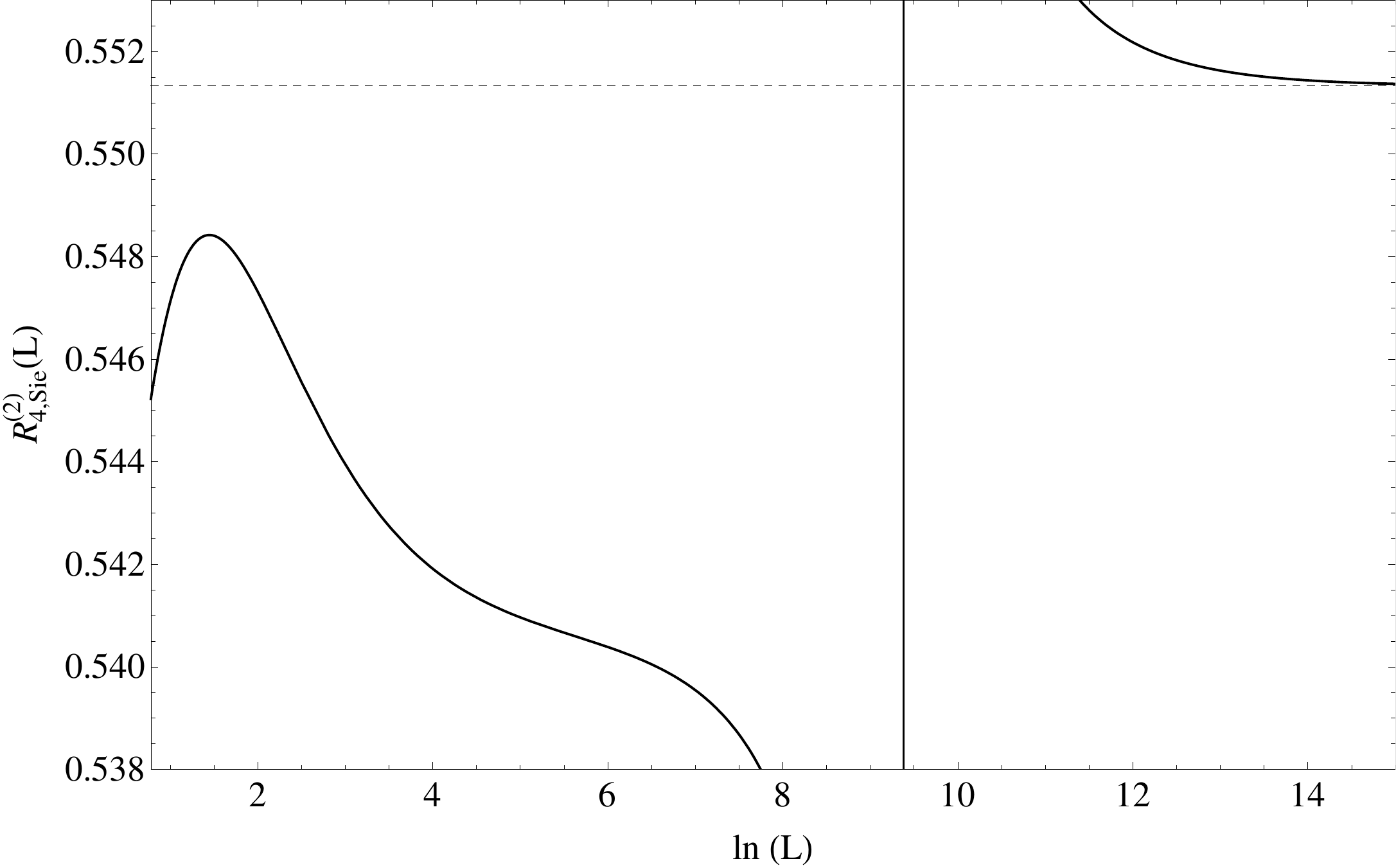}
\caption{Plot of $R^{(2)}_{4,sie}(L)$ for the tachyon vev as a
function of level in Siegel gauge. The dashed line represents the
conjectured value $\sqrt{3}/\pi$.} \label{THfig2}
\end{figure}

By adding the two more extra data points given in the  fourth
column of table \ref{resultsA2} (the ones at $L=12$ and $L=14$),
we can obtain the interpolating function of order
$R^{(2)}_{4,Sie}(L)$. We expect that the behavior of the function
$R^{(2)}_{4,Sie}(L)$ will not be much different than the function
$R^{(2)}_{3,Sie}(L)$. For instance, the point where the tachyon
vev starts to increase obtained by using $R^{(2)}_{4,Sie}(L)$
should be close to the one obtained by using $R^{(2)}_{3,Sie}(L)$.
However, by analyzing $R^{(2)}_{4,Sie}(L)$ we observe that there
is no local minimum where the tachyon vev starts to increase,
moreover there is a point of discontinuity near $L \sim 11908$.
Figure \ref{THfig2} clearly illustrates our point.

From the above results and by computing higher interpolating
functions (with $n>4$), we conclude that the set of functions
$R^{(2)}_{n,Sie}(L)$ do not seem to converge to some smooth
function in the limit case $n \rightarrow \infty$. This bad
behavior indicates that scenario $S_2$ is not the correct one.

\begin{figure}[h]
\centering
\includegraphics[width=5.5in,height=85mm]{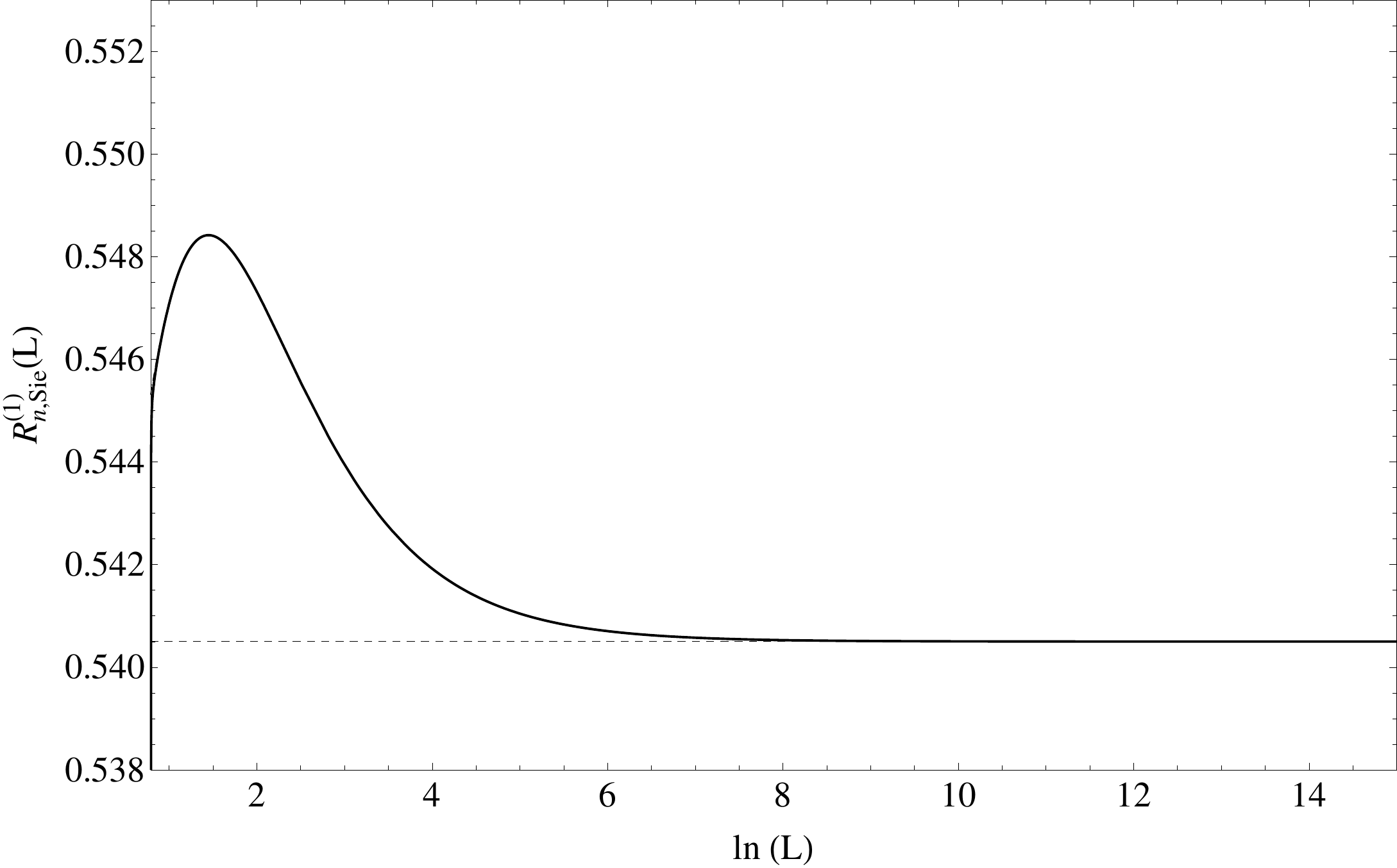}
\caption{Plot of the functions $R^{(1)}_{n,Sie}(L)$ for the
tachyon vev as a function of the level in Siegel gauge. The dashed
line represents the asymptotic value $0.5405$ obtained in
reference \cite{Gaiotto:2002wy}. The curves for $n = 3, 4, 5, 6,
7$ appear superimposed in the figure.} \label{THfig3}
\end{figure}

By performing similar analysis for the case of the interpolating
functions $R^{(1)}_{n,Sie}(L)$, namely, for scenario $S_1$ we get
a nice behavior that is illustrated in figure \ref{THfig3}, the
functions $R^{(1)}_{n,Sie}(L)$ converge to a smooth function in
the limit case when $n \rightarrow \infty$. This is a clear
indication that scenario $S_1$ is the right one. In the following
subsection, we are going to analyze the case of Schnabl gauge.

\subsection{Schnabl gauge}
The results for the tachyon vev in Schnabl gauge obtained from
$(L,3L)$ level-truncation computations are shown in table
\ref{TabSch1}. Note that as in the case of Siegel gauge, the value
of the tachyon vev has a maximum value near level $L=4$ and then
starts to decrease.

Since in the case of Schnabl gauge the analytic (asymptotic) value
of the tachyon vev is known \cite{Schnabl:2005gv}, the two
scenarios $S_1$ and $S_2$ should be the same. Let us compute the
interpolating functions $R^{(2)}_{n,Sch}(L)$ corresponding to
scenario $S_2$.

We use a rational function in $L$ to interpolate the data for the
tachyon vev shown in table \ref{TabSch1} together with the
asymptotic point at $L \rightarrow \infty$
\begin{align}
\label{schta1}R^{(2)}_{3,Sch}(L)= \frac{a_0 + a_1 L + a_2 L^2 +
a_3 L^3}{1 + b_1 L + b_2 L^2 + b_3 L^3}.
\end{align}
Since we have seven data points (which include the point at
infinity), we have set $n=3$. As usual, to obtain the seven
unknown coefficients $a_i$ and $b_i$, we require that the data
points in table \ref{TabSch1} coincide with the rational function
(\ref{schta1}) evaluated at the known values of the direct
$(L,3L)$ level-truncation computations for $L=0,2,4,6,7,10$ and
the asymptotic one $L\rightarrow\infty$ (the analytic result). For
instance, at level $L=0$, we obtain
\begin{align}
\label{schta2} R^{(2)}_{3,Sch}(L=0) = a_0 = 0.456177990470,
\end{align}
while using the asymptotic value, we have the following equation
\begin{align}
\label{schta3} \lim_{ L \rightarrow \infty} R^{(2)}_3(L) =
\frac{a_3}{b_3} = 0.553465566934.
\end{align}
In this way, we get a system of seven equations for the
coefficients $a_i$ and $b_i$ which can be easily solved. Once
these coefficients are known, the next step is to analyze the
rational function $R^{(2)}_{3,Sch}(L)$.

\begin{figure}[h]
\centering
\includegraphics[width=5.5in,height=85mm]{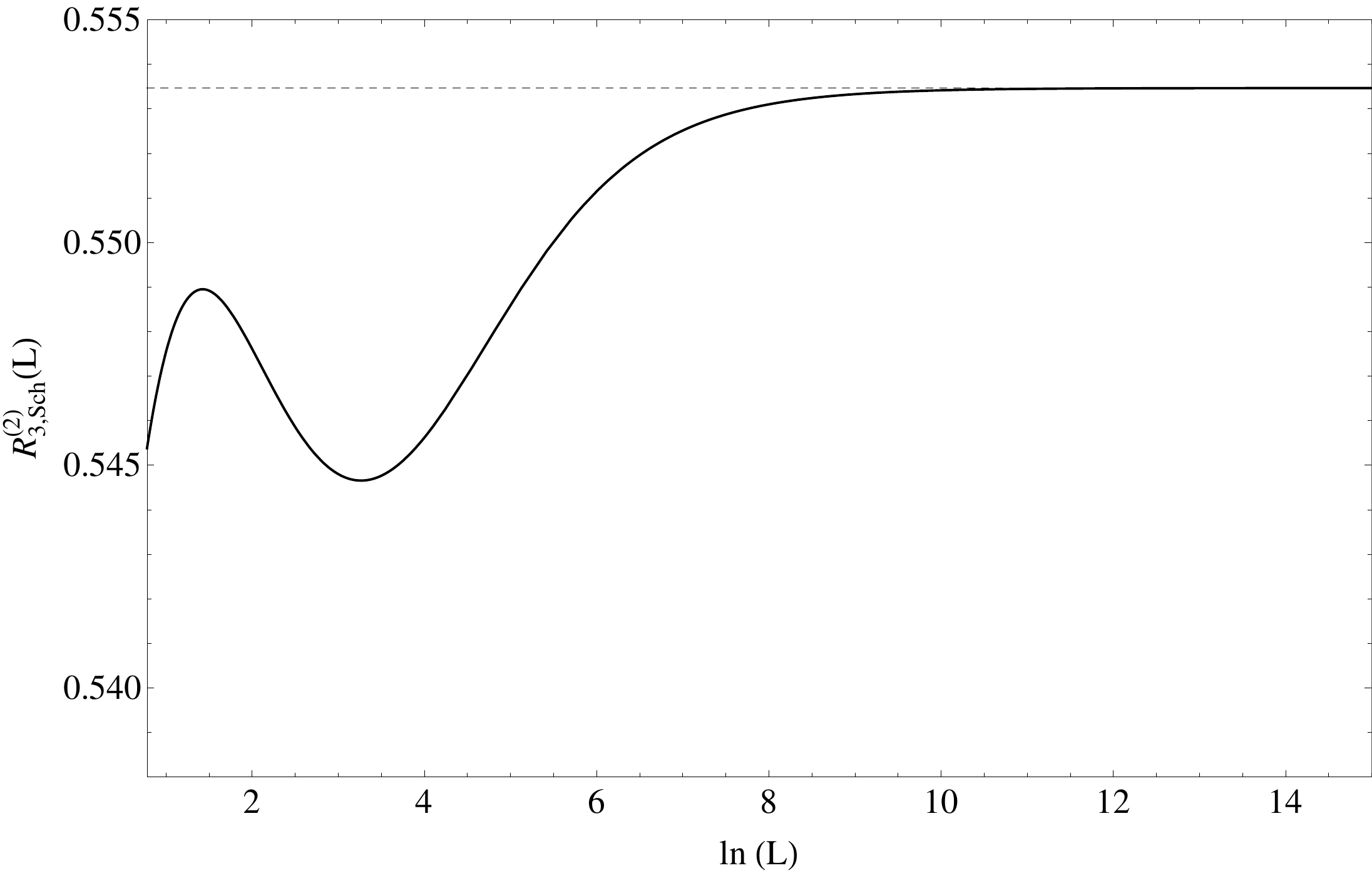}
\caption{Plot of the function $R^{(2)}_{3,Sch}(L)$ shown as the
continuous line. The dashed line represents the asymptotic value
$0.5534$ obtained in reference \cite{Schnabl:2005gv}.}
\label{THfig4}
\end{figure}

Let us evaluate $R^{(2)}_{3,Sch}(L)$ for values of $L>10$ between
$12$ to $26$, the results are shown in table \ref{resultsA5}. Note
that for levels close to $L=24$, the value of the tachyon vev
seems to stop decreasing. At levels, between $L=24$ and $L=26$ the
value of the tachyon vev is almost the same, this indicates that
we are close to a point where we have reached a local minimum. In
figure \ref{THfig4}, we show the plot of $R^{(2)}_{3,Sch}(L)$, as
we can see, at level close to $L=26$, the tachyon vev starts
increasing and then approaches asymptotically the analytic value
shown as the dashed line.

\begin{table}[ht]
\caption{The rational function $R^{(2)}_{3,Sch}(L)$ for some
values of $L>10$.} \centering
\begin{tabular}{|c|c|}
\hline $L$ & $R^{(2)}_{3,Sch}(L)$  \\
\hline
12 & 0.545904273712  \\
\hline 14 & 0.545468779861  \\ \hline
16 & 0.545160525135 \\
\hline 18 & 0.544947247026  \\ \hline 20 & 0.544805165981 \\
\hline
22 & 0.544716903893  \\
\hline 24 & 0.544669683450 \\
\hline
26 & 0.544654007673  \\
\hline
\end{tabular}
\label{resultsA5}
\end{table}

\begin{figure}[h]
\centering
\includegraphics[width=5.5in,height=85mm]{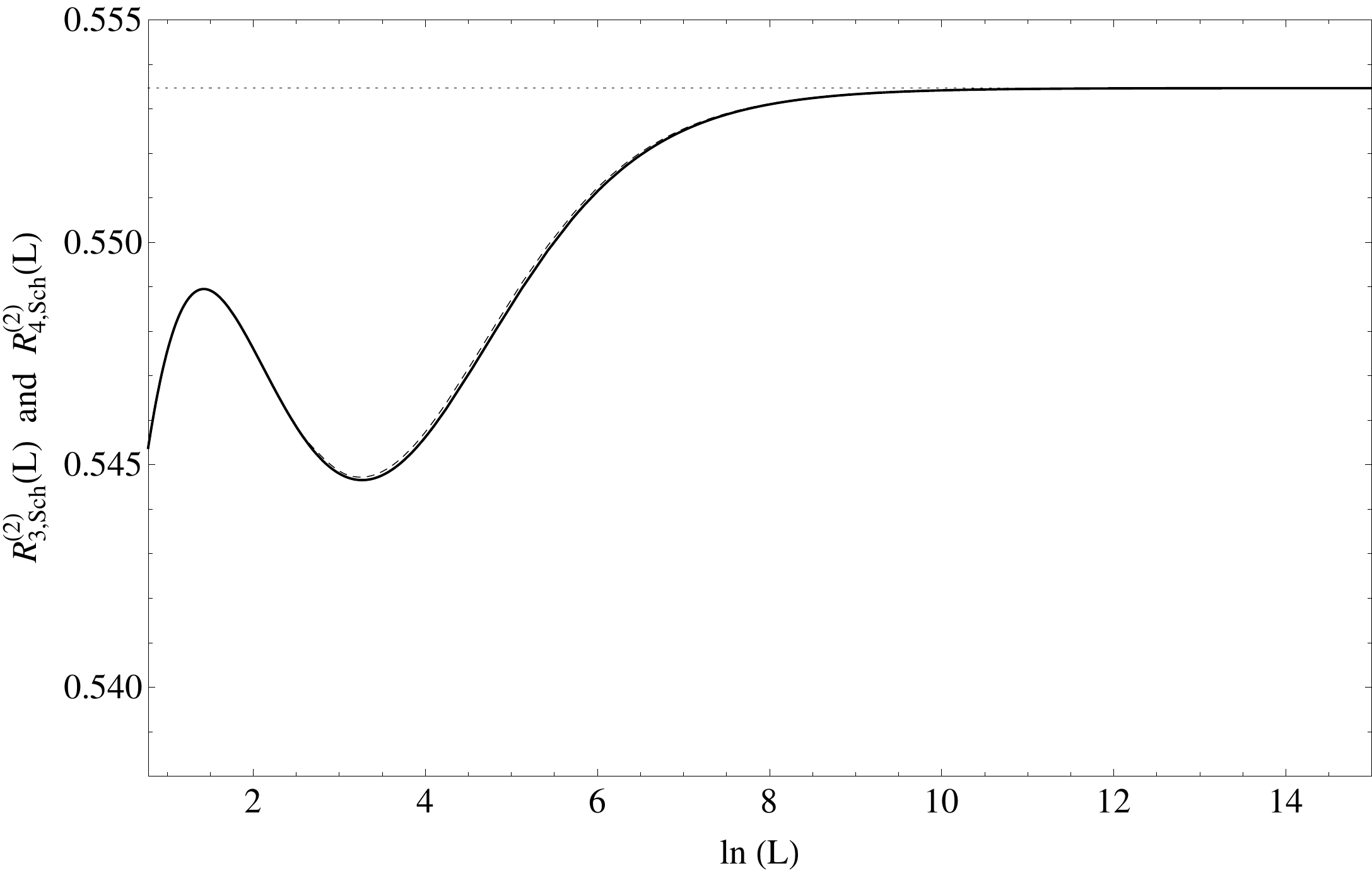}
\caption{Plot of the function $R^{(2)}_{3,Sch}(L)$ shown as the
continuous line compared with the function $R^{(2)}_{4,Sch}(L)$
represented as the dashed line for the tachyon vev as a function
of the level in Schnabl gauge. The dotted line represents the
asymptotic value $0.5534$ obtained in reference
\cite{Schnabl:2005gv}.} \label{THfig5}
\end{figure}

By adding two more extra data points obtained by direct $(L,3L)$
level-truncation computations for $L=12$ and $L=14$, we can derive
the interpolating function of order $R^{(2)}_{4,Sch}(L)$. We
expect that the behavior of this function $R^{(2)}_{4,Sch}(L)$
will not be much different from the function $R^{(2)}_{3,Sch}(L)$,
for instance, the point where the tachyon vev starts to increase
obtained by using $R^{(2)}_{4,Sch}(L)$ should be close to the one
obtained by using $R^{(2)}_{3,Sch}(L)$. In fact, this behavior is
observed as shown in figure \ref{THfig5}. We expect that when
$n\rightarrow\infty$ the set of functions $R^{(2)}_{n,Sch}(L)$
converges to some smooth function.

To be honest, since we have not computed the values of the tachyon
vev for $L=12$ and $L=14$ by means of direct $(L,3L)$
level-truncation computations, to derive the interpolating
function $R^{(2)}_{4,Sch}(L)$ we have used the data shown in the
second column of table \ref{resultsA5} for $L=12$ and $L=14$ with
four digits of precision, namely, the values $0.5459$ and $0.5454$
respectively. We hope that the actual $(L,3L)$ level-truncation
computation will confirm these values for the tachyon vev.

The above results suggest that at level close to $L \sim 26$, the
value of the tachyon vev starts to increase until reaching the
asymptotic value of $0.5534$. Figure \ref{THfig5} illustrates this
point. We leave as a future research project, the test of the
validity of this prediction by means of direct $(L,3L)$
level-truncation computations for levels $L>10$.

\section{Summary and discussion}
Using either Siegel or Schnabl gauge, we have constructed the
effective tachyon potential and analyzed its branch structure by
means of Virasoro $L_0$ level-truncation computations. It would be
interesting to extend this analysis to higher levels and to probe
the physical branch in a much larger region. We should use a
different gauge such that its region of validity must be greater
than Siegel or Schnabl gauges, for instance, we could explore the
so-called linear $b$-gauges studied in reference
\cite{Kiermaier:2007jg}.

Selecting the physical branch of the effective tachyon potential
in Schnabl gauge, namely, the branch that connects the
perturbative with non-perturbative vacuum and computing its local
minimum, we have derived data points for the energy as well as for
the tachyon vev.

Regarding the data for the energy obtained by direct $(L,3L)$
level-truncation computations, we have observed that at level
$L=6$ the energy overshoots the expected analytical answer of
$-1$, and appears to further decrease at higher levels. This
result indicates that the approach of the energy to $-1$ as $L
\rightarrow \infty$ is non-monotonic. By applying two kind of
extrapolation techniques to the level-truncation data for $L\leq
10$ to estimate the vacuum energies even for $L > 10$, we have
predicted that the energy reaches a minimum value at $L \sim 12$,
and then turns back to approach asymptotically $-1$ as $L
\rightarrow \infty$.

Regarding the data for the tachyon vev, we have found that
starting at level $L=4$, the value of the tachyon vev decreases.
To reach the analytical value of $0.553465$, we have noted that
there should be some higher value of $L>4$ such that the value of
the tachyon vev stops decreasing and then starts increasing until
reaching this analytical value. We have predicted that for $L>4$
the tachyon vev reaches a minimum value for $L \sim 26$, and then
turns back to approach asymptotically the expected analytical
result.

By explicit $(L,3L)$ level-truncation computations with $L>10$, it
would be interesting to confirm the above predicted results. Since
the minimum value of the energy data should happen at $L \sim 12$,
the direct $(L,3L)$ level-truncation computations must be
performed, at least, up to level $L=14$. While in the case of the
tachyon vev data, to confirm the existence of a minimum value
close to level $L \sim 26$, we will need to perform the
calculations, at least, up to level $L=28$. These issues will be
the subject of a future research project.

Finally, since the modified cubic superstring field theory
\cite{Arefeva:1989cp} as well as Berkovits superstring field
theory \cite{Berkovits:1995ab} are based on Witten's associative
star product of open bosonic string field theory, using rather
generic gauge conditions, our results can be naturally extended to
analyze numerical solutions in the context of these open
superstring field theories.

\section*{Acknowledgements}
EAA would like to thank Ted Erler and Isao Kishimoto for useful
discussions. The research of AFS is supported by CAPES grant
1675676. The research of RS is supported by CAPES grant 1627292.




\end{document}